\documentclass[dvipsnames,acmsmall]{acmart}

\AtBeginDocument{%
  \providecommand\BibTeX{{%
    \normalfont B\kern-0.5em{\scshape i\kern-0.25em b}\kern-0.8em\TeX}}}

\setcopyright{acmlicensed}
\copyrightyear{2025}
\acmYear{2025}
\acmDOI{XXXXXXX.XXXXXXX}

\acmConference[CSCW '25]{The 28th ACM SIGCHI Conference on Computer-Supported Cooperative Work \& Social Computing}{October 18--22,
  2018}{Bergen, Norway}
\acmISBN{978-1-4503-XXXX-X/18/06}

\usepackage{subcaption}
\usepackage{graphicx}
\usepackage{colortbl}
\usepackage{xcolor}
\usepackage{enumitem}
\usepackage{courier}

\newcommand{\revision}{\textcolor{black}}

\begin{document}

\title[``I'm Petting the Laptop, Which Has You Inside It'']{``I'm Petting the Laptop, Which Has You Inside It'': \\Reflecting on Lived Experiences of Online Friendship}

\author{Seraphina Yong}
\affiliation{%
  \institution{University of Minnesota}
  \city{Minneapolis, MN}
  \country{USA}}
\email{yong0021@umn.edu}

\author{Ashlee Milton}
\affiliation{%
  \institution{University of Minnesota}
  \city{Minneapolis, MN}
  \country{USA}}
\email{milto064@umn.edu}

\author{Evan Suma Rosenberg}
\affiliation{%
  \institution{University of Minnesota}
  \city{Minneapolis, MN}
  \country{USA}}
\email{suma@umn.edu}

\author{Stevie Chancellor}
\affiliation{%
  \institution{University of Minnesota}
  \city{Minneapolis, MN}
  \country{USA}}
\email{steviec@umn.edu}

\author{Svetlana Yarosh}
\affiliation{%
  \institution{University of Minnesota}
  \city{Minneapolis, MN}
  \country{USA}}
\email{lana@umn.edu}

\renewcommand{\shortauthors}{Anonymous, et al.}

\begin{abstract}
Online(-only) friendships have become increasingly common in daily lives post-COVID despite debates around their mental health benefits and equivalence to ``real'' relationships.
Previous research has reflected a need to understand how online friends engage beyond individual platforms, and the lack of platform-agnostic inquiry limits our ability to fully understand the dynamics of online friendship.
We employed an activity-grounded analysis of 25 interviews on lived experiences of close online friendship spanning multiple years. Our findings present unique challenges and strategies in online friendships, such as stigma from real-life circles, an ambivalent relationship with online communities, and counter-theoretical reappropriations of communication technology.
This study contributes to HCI research in online communities and social interface design by refocusing prior impressions of strong vs. weak-ties in online social spaces and foregrounding time-stable interactions in design for relationship maintenance through technology.
Our work also promotes critical reflection on biased perspectives towards technology-mediated practices and consideration of online friends as an invisible marginalized community.
\end{abstract}



\received{20 February 2007}
\received[revised]{12 March 2009}
\received[accepted]{5 June 2009}

\maketitle

\section{Introduction}
\begin{quote}
    \textit{``True friends are never apart. \\Maybe in distance but never in heart."} - Helen Keller
\end{quote}
We are in the middle of a friendship epidemic -- people are struggling to sustain close friendships \cite{armstrong_friendships_2022}. Surveys have shown an increase in the amount of Americans with no close friends from 2 to 12\% since 1990 \cite{cox_state_2021}, and over half of UK residents report difficulty making any friends at all~\cite{dinic_yougov_2021}. 
Yet, digitalization has facilitated the rise of making friends online~\cite{chayko_superconnected_2020}.
Since 2021, nearly 40\% of Americans and over a quarter of Britons have made \textbf{online friends} (also known as online-only, or `internet friends,'). These are friends\revision{\footnote{Friendship \cite{rubin_handbook_2009} involves mutual liking between individuals, separate from (yet potentially coexisting with) romantic/sexual attraction. This reciprocal liking sustains friendships, unlike the explicit commitment structures in romantic partnerships. Friendship dynamics can evolve over time and vary individually (e.g., romantic relationships can develop from friendships).}} whom one has met \revision{online} \textit{and} primarily contacts through online platforms \cite{cox_state_2021,dinic_yougov_2021}.
Online friendships hold significance for people who lack companionship in their physical surroundings, especially those who face stigma, have health conditions, or have unique interests \cite{mcinroy_building_2020,sukthankar_good_2018,ringland_will_2016,tian_social_2013,ghosh_i_2023}.
However, public impressions that ``real'' intimacy can only occur physically paint online friendship as a substandard form of social connection, reflecting the longstanding stigma that it carries \cite{chayka_lets_2015,amichai-hamburger_friendship_2013}.

Inconclusive research in this area and our limited understanding of the mechanisms behind online friendship make it difficult to judge the true quality or purpose of online friendship.
Multiple post-COVID studies have documented the rise of online friendships but contain \revision{contrasting} results on whether online friendship can facilitate connection \cite{cairns_covid-19_2020,lariviere-bastien_childrens_2022,towner_virtual_2022,patulny_beware_2022,lee_friendship_2024}, mirroring older studies \revision{whose} results when comparing online friendships to that of `real-life' friendships \revision{also conflict with one another} \cite{chan_comparison_2004,antheunis_quality_2012}.
In-depth studies of online friendship tend to be conducted across isolated and disparate platforms and interaction contexts \cite{scott_social_2022,utz_social_2000,lai_online_2020,sheng_virtual_2020,sukthankar_good_2018}, despite discussing the need for more holistic approaches towards understanding experiences of online friendship \cite{sukthankar_good_2018,sheng_virtual_2020,tang_development_2010}.
\revision{The trajectories of online friendship not only vary based on context, but also evolve and shift over time.}
The urgency of preserving high-quality friendship in our current era \cite{ren_building_2012,sassenberg_common_2002,davis_friendship_1982,fuhrman_behavior_2009} warrants deeper clarification of the nature and value of online friendships, which are distinct from offline-created (IRL) friendships and other types of relationships \cite{antheunis_quality_2012}.

Development of close, quality friendship is a long-term process \cite{hays_development_1984,fehr_friendship_1996,oswald_friendship_2004}, and we propose that understanding core identifying qualities of online friendship requires integrating these fragmented insights from prior work into a \textit{developmental} perspective of online friendship: as an evolving \textit{process} grounded in the entire sociotechnical ecosystem.
We define \textit{sociotechnical ecosystem} as the full technological and social environment of individuals involved in an online friendship, including all affordances for social contact and influencing actors in both online and physical settings.
The real-world construct of online friendship deeply intersects with key areas of HCI, such as social connection in online communities \cite{arguello_talk_2006,lampe_motivations_2010} and technology-mediated relationship maintenance \cite{vetere_mediating_2005}. However, we need a better-contextualized understanding of online friendships in their unique setting to expand our insights in these relevant areas.
For example, what actual sociotechnical conditions deepen online friendships or break them apart? Why might friends choose to move to a different platform, or use more than one at the same time?

We adopt a \textit{process-oriented} lens to investigate the context of online friendship development and pose the following research questions:

\begin{quote}
\textbf{RQ1:} What are the  characteristics of online friendship \textit{development}?\\
\textbf{RQ2:} How does online friendship \textit{development} function within its sociotechnical ecosystem?
\end{quote}
We interviewed 25 people about the close online-only friendships they developed and maintained across years over a diverse selection of platforms, and used Charmaz's grounded theory approach \cite{charmaz_constructing_2006} to analyze the online friendship process as an activity system \cite{seaman_adopting_2008}.

Our study revealed online friends' complicated relationship with their surrounding online and physical worlds, highlighting distinctive approaches they took to preserve their virtual existence.
Contradicting a `better or worse' perspective of online friendship, we saw that online friendships operated on norms of interaction different from those in real-life, contributing to pressure and stigma placed on friendships due to lack of understanding.
Online friends worked to achieve `temporal constancy,' a permanent sense of mutual presence, by creatively re-appropriating communication options (e.g., voice calls) to simulate physical activities with one another and capitalizing on interactions which persist over time to develop shared history.
They experienced a complex relationship with social environments in online platforms (e.g., communities, teams, or message groups), which supplied momentum for maintaining consistent contact but also propagated group dynamics harmful to interpersonal bonding.

In summary, our contributions are:
\begin{itemize}
    \setlength\itemsep{0.5em}
    \item Rich descriptions of online friendship development through an interview study collecting lived experiences across a diversified selection of platforms, activities, and communities.
    \item Grounded insights for the online friendship design space by adopting an \textit{activity-based} conceptualization which identifies clear points of connection between elements of the temporal friendship-building process and sociotechnical features of different online platforms.
    \item An HCI research agenda to substantively account for online friendship development in sociotechnical systems: \textbf{addressing stigma} and its influence on technology-mediated friendship, acknowledging tensions between \textbf{strong and weak ties} in online communities, and designing for \textbf{time-stable interaction} in technology-mediated relationships.
\end{itemize}

\section{Background}
We contextualize this work by describing the overlapping scope of HCI research broadly covering social connection in online communities, as well as adjacent work focusing on online-only friendships themselves.
We follow with a brief introduction of Activity Theory and the history of its use in inquiries of online socialization.

\subsection{Social Support and Connection in Online Communities}
\label{subsec:socialcommunities}
HCI has amply explored the role of social connection in online communities \cite{arguello_talk_2006,lampe_motivations_2010}.
Social support through peer-to-peer connection is present in a diverse range of online platforms, including forum spaces, social virtual reality (VR), and massively multiplayer online environments (MMOs) \cite{prescott_peer_2017,sukthankar_good_2018,deighan_social_2023,li_we_2023}. 
Specific online spaces can lend significant social support to vulnerable and stigmatized groups, such as neurodiverse youth and gender minorities.
Work by Ringland et al. demonstrated autistic youth's self-conceptualization of sociality through digitally-embodied play in Minecraft as a safe space \cite{ringland_will_2016,ringland_place_2019}.
Gender diverse youth have also developed their identities and built resilience through mutual support with similar peers through online safe spaces \cite{austin_its_2020,han_what_2019,hillier_internet_2012}. Some of these spaces take the form of fandom communities which enable shared world-building based on personally-resonant concepts \cite{mcinroy_building_2020,ghosh_i_2023}, or spaces in social VR which afford embodied identity expression and rapport building \cite{acena_my_2021,li_we_2023}. 
More broadly, people also share experiences to gain mutual understanding and form friendships on online mental health forums \cite{prescott_peer_2017}, online health communities (OHCs) \cite{gatos_how_2021}, and more recently social VR, which was perceived as a safe space to practice social interaction and form connections during COVID-19 \cite{deighan_social_2023}. 

Online gaming environments such as MMOs have been particularly studied as \textit{``third'' places} which support different forms of community-building social activities that approximate face-to-face interactions in a physical space \cite{steinkuehler_where_2006,moore_3d_2009}. However, findings indicate that task-based interactions and informal social affordances offered more support for casual socializing and less for the development of deeper emotional bonds between users, such as friendship.
Instead, Ramirez et. al \cite{sukthankar_good_2018} identified \textit{backchannel communication} (players using out-of-game media platforms to communicate) as a core component of maintaining close online-only friendships created in the MMO. These strong ties were distinct from users' casual friends and became independent from the original MMO.
The authors note: \textit{``The contents of backchannel communication and the everyday experiences of multisited engagement remain underexplored,''} indicating the need to understand online friendship at a greater scope.
Works analyzing friendship on other community spaces (e.g. discussion forums or livestream platforms) \cite{tang_development_2010,sheng_virtual_2020} have also echoed this finding and sentiment that research on online friendship needs to account more fully for the larger ecosystem.

The above findings show that forming online social bonds can benefit the mental health of those who are seeking connection rather than exacerbating loneliness \cite{stuart_online_2021,deighan_social_2023,amichai-hamburger_loneliness_2003,morahan-martin_loneliness_2003}, but they also identify a need to understand the full ecosystem of close online friendship. 
For online friends, developing the complex routine of behaviors required for \textit{maintaining} quality and commitment in a friendship over time \cite{hays_development_1984,fehr_friendship_1996,oswald_friendship_2004} depends heavily on the sociotechnical affordances available to them --- leading our work to focus on a broad scope of platform contexts for friendship development.

\subsection{Online-only `Internet' Friendships}
\label{subsec:internetfriends}
Research on internet friendships to-date started from Richard Walther’s work in the early 1990s, which coincided with the rise of the World Wide Web and introduced the argument that meaningful personal relationships can be developed even through fully computer-mediated communication (CMC) \cite{walther_interpersonal_1992}. 
In this work, Walther also introduced \textit{social information processing} (SIP) theory \cite{walther_social_2008} to the field of CMC, which asserts that people develop relationships by adapting to the contextual cues of their social environment \cite{lampe_familiar_2007,walther_interpersonal_1992,whitty_agesexlocation_2001}.
Unrestricted to an absolute model of media richness where face-to-face contact is superior to CMC \cite{daft_organizational_1986}, people shape their communication patterns based on the cues present. Rather than on the face-value of the sociotechnical environment, relationship quality depends more on how users manage their time spent interacting with one another and how they choose to apply available cues \cite{walther_interpersonal_1992}.
This perspective is also supported by Walther's \textit{hyperpersonal model} \cite{walther_computer-mediated_1996}, which describes how CMC can surpass in-person communication in facilitating closeness, as well as Bonnie Nardi's work on \textit{dimensions of connection}, which emphasizes the impact of interactions across time on relational quality \cite{nardi_beyond_2005}.

Since then, empirical work on internet friendships in particular has explored broad social factors which impact relationship quality.
Self-disclosure is commonly studied due to its crucial role in developing intimate relationships \cite{altman_social_1973} and relevance to the \textit{hyperpersonal model} (self-disclosure is easier online due to the lack of perceived physical boundaries or IRL social norms which may restrict expression \cite{walther_computer-mediated_1996,baym_personal_2010,briggle_real_2008}).
Research on cultural and individual differences, such as Yum and Hara's cross-cultural survey on internet friendship development among Americans, South Koreans, and Japanese individuals \cite{yum_computer-mediated_2005} showed that self-disclosure improved general relationship quality for all cultures, but only increased sense of trust in Americans.
Other works have found that individuals with social anxiety self-disclose and form higher quality friendships more easily with people who they meet online \cite{tian_social_2013,amichai-hamburger_friendship_2013,morahan-martin_loneliness_2003}.
However, general comparisons between the quality of online and offline friendships have mixed and inconclusive results: some work shows that differences in their quality decrease over time \cite{chan_comparison_2004}, while others show that those differences remain significant \cite{antheunis_quality_2012}. Recent post-COVID work adds context by showing that individuals with primarily offline friendships were less satisfied with online friendships, compared to those with online friendship experience prior to the pandemic \cite{scott_social_2022}.
These mixed findings suggest the need for a deeper understanding of what sociotechnical factors mediate online friendship as a whole.

Research which examines the online friendship development process in more depth has fallen primarily within the scope of social gaming, covering contexts such as multiplayer game environments, livestreaming communities, and video game console networks \cite{utz_social_2000,lai_online_2020,sheng_virtual_2020,sukthankar_good_2018,freeman_revisiting_2016}.
Motivation to maintain ongoing activities with the other person heavily mediated friendships' closeness and success. \textit{Backchanneling} (e.g. instant messaging, see \ref{subsec:socialcommunities}) provided private spaces for self-disclosure and personal interactions, and reduced reliance on eventually-discontinued platforms \cite{lai_online_2020,sheng_virtual_2020,sukthankar_good_2018}.
Although shared common interests and activities were strong factors for friendship formation \cite{lai_online_2020,sheng_virtual_2020}, players who were too task-focused did not engage enough socially to build friendships \cite{utz_social_2000}.
This complex interplay between personal motivations, activity context, and platform migration just within the social gaming context further illustrates the need to understand how these processes extend to online friendships which are known to form in other spaces.

Past work has shed light on various elements of online friendship, but \revision{lack a process-oriented understanding, which can contribute a more holistic perspective.}
Our work extends this body of research through an activity-focused lens to provide actionable insights on how elements of the online friendship development process are grounded in features of the sociotechnical ecosystem.

\subsection{Online Socialization Through an Activity System Lens}
\label{subsec:atintro}
Activity theory (AT), also known as \textit{cultural-historical activity theory}, provides an integrating framework for HCI researchers to account for the influences of social and cultural context when studying technology-mediated activity \cite{kaptelinin_computer-mediated_1995,leontev_activity_1978}. We will use {\small\texttt{this notation}} to refer to elements and meta-features of the activity system in this work.

Stemming from Vygotsky's sociocultural theory of human development \cite{kozulin_vygotskys_1990}, the AT framework accounts for not only the user ({\small \texttt{subject}}) and technologies/affordances ({\small \texttt{tools}}) of immediate interest but also their connections to the social environment ({\small \texttt{community}}), as well as norms ({\small \texttt{rules}}) and community member impacts ({\small \texttt{division of labor}}) imposed on the problem space ({\small \texttt{object}})---which, in our case, is online friendship development.
These six {\small \texttt{elements}} comprise the \textbf{activity system} (see Figure~\ref{fig:activitydiagram} from our Findings), which continually morphs as structural tensions accumulate and change over time.
In Engestr\"{o}m's extension of AT, such tensions are referred to as {\small \texttt{contradictions}} in the system and drive the development of the historically-grounded {\small \texttt{activity}} as an evolving social process.
{\small \texttt{Subjects}} adapt new actions as responses to these situations or emerging opportunities to transform the {\small \texttt{object}} into the desired {\small \texttt{outcome}} \cite{engestrom_learning_2014,kuutti_activity_1995}.
For ease of understanding, in this paper we will refer to such {\small \texttt{contradictions}} as {\small \textbf{\texttt{tensions}}}.

Prior work hints at the relevance of AT to the topic of online friendship.
The cultural context of different online social platforms influences norms of use and behavioral motivations for forming online friendships (see \ref{subsec:socialcommunities} and \ref{subsec:internetfriends}). Existing work has also identified potential {\small \texttt{tensions}} that the sociotechnical ecosystem presents for online friendship, such as balancing between task focus and socialization \cite{utz_social_2000} as well as the complexity of \textit{hyperpersonal} communication \cite{walther_computer-mediated_1996} in multi-platform engagement \cite{lai_online_2020}.
Prior work in HCI has also applied AT to personal connection and aesthetic experience in virtual worlds, such as Nardi's seminal account of social roleplay in World of Warcraft \cite{nardi_my_2010}, and her work on computer-mediated connections mediated by a history of activity \cite{nardi_beyond_2005}.

To provide an understanding of online friendships that is both sensitive to sociotechnical context and more broadly applicable in HCI, we use a grounded-theoretical approach to construct the elements of online friendship development as an activity system spanning the full sociotechnical ecosystem available to online friends. Our analysis identifies key {\small \texttt{tensions}} encountered in the {\small \texttt{transformation}} process, or development, of online friendship (see more in Section \ref{subsec:analysis}).



\section{Methods}
We conducted 25 semi-structured interviews where we asked participants about their close friendship with someone they met online.
We used theoretically-informed sampling \revision{(see Section \ref{subsec:participants})} to conduct semi-structured interviews with a process-oriented elicitation and questions focusing on friendship development. We used a grounded-theoretical approach to analyze our interviews and constructed the elements of online friendship development as an activity system of online technologies. The following sections describe our participants, interview procedures, and analysis methods. 
\revision{This study, which recruits teens 16 and older, underwent full board review by the Institutional Review Board at the author's institution and was approved for waiving parental consent after some discussion. This was due to the risk of the interview being no greater than the risk of speaking about these online friendships in a non-research setting, in addition to known problems in enforcing parental control over teen disclosure of online social activities \cite{wisniewski_parents_2017,wisniewski_parental_2017}.}

\subsection{Participants}
\label{subsec:participants}
We recruited across multiple platforms, mainly targeting Discord servers and subReddits surrounding ideas or topics highlighted in prior work on online friendships (e.g., online gaming, livestreaming, fandoms, social VR, etc.)\cite{utz_social_2000,lai_online_2020,sheng_virtual_2020,sukthankar_good_2018,mcinroy_building_2020,deighan_social_2023}. 
Interested individuals filled out an online screener checking for eligibility: 16 years of age or older and had at least one close online-only friendship \revision{(defined based on each individual's own experiences and understanding of friendship \cite{rubin_handbook_2009})}. If they were eligible, they proceeded to the consent form, demographic, and online friendship history survey.
The researchers used theoretically-informed sampling \cite{seaman_adopting_2008} using participant-provided information to select interviewees among eligible participants. The interviewee pool was selected to balance representation among the different types of platforms and communities mentioned in prior work on online friendships.

All interviews were conducted and recorded via Zoom from November 2023 to March 2024 with participant consent. Out of 118 people who signed up, 67 were eligible for the study. 
\revision{We selected interviewees from the eligible pool to balance representation of known platforms, communities, and demographics.}
The first author \revision{scheduled and} conducted interviews until theoretical saturation \cite{charmaz_constructing_2006} was reached at the agreement of the first 2 authors, and no new themes emerged for several interviews, resulting in 25 interviews.
\revision{Participants were compensated with a \$25 Amazon e-gift card upon completion of the interview. The average length of each interview was 90 minutes.}

The demographic information of our participants can be found in \autoref{Tab:Demographic}. Since we allowed for self-identification and multiple selections, counts of demographic characteristics do not necessarily sum to the number of participants. Our participants ranged in age from 16 to 37 (M=25.4) and were predominantly based in the United States (N=18). The gender of our participants is slightly skewed to women (N=13), and our largest pool of participant ethnicities is white (N=8).

\begin{table}[]
\centering
\resizebox{0.7\linewidth}{!}{%
\begin{tabular}{|r|l|r|l|}
\hline
\multicolumn{1}{|l|}{\textbf{Demographic Variables}} &
  \textbf{N} &
  \multicolumn{1}{l|}{\textbf{Demographic Variables}} &
  \textbf{N} \\ \hline
\rowcolor[HTML]{C0C0C0} 
\textbf{Age}          &    & \textbf{Ethnicity}        &    \\
16-20         & 8  & Asian            & 5  \\
21-25         & 5  & Black            & 2  \\
26-30         & 7  & Indian           & 1  \\
31-35         & 4  & Hispanic         & 3  \\
36-40         & 1  & Khazakstani      & 1  \\
\cellcolor[HTML]{C0C0C0}\textbf{Gender}        &  \cellcolor[HTML]{C0C0C0}  & Native American  & 1  \\
Man           & 12 & Pacific Islander & 1  \\
Woman         & 10 & White            & 12 \\
Non-Binary    & 3  & \cellcolor[HTML]{C0C0C0} \textbf{Location}         &  \cellcolor[HTML]{C0C0C0}  \\
\cellcolor[HTML]{C0C0C0} \textbf{Education}     &  \cellcolor[HTML]{C0C0C0}  & Australia        & 1  \\
High School   & 6  & Canada           & 1  \\
Undergraduate & 12 & Germany          & 1  \\
Graduate      & 7  & Khazakstan       & 1  \\
              &    & United Kingdom   & 3  \\
              &    & United States    & 18 \\ \hline
\end{tabular}%
}
\caption{Demographic information}
\label{Tab:Demographic}
\end{table}

We also provide information about participants' online friendships in \autoref{Tab:Meet}. The duration of the close online friendship described by each participant ranged from just under a year to 14 years, with an average of around 5 years. Our participants met their online friends in a variety of ways, while most centered around gaming, fandom, or building a community around a shared interest.
Different platforms were used during friendship, but Discord was used by the vast majority and about half of the participants played an online game with their friend.

\begin{table}[]
\centering
\resizebox{\linewidth}{!}{%
\begin{tabular}{|p{.075\linewidth}|p{.125\linewidth}|p{.4\linewidth}|p{.4\linewidth}|}
\hline
\rowcolor[HTML]{C0C0C0} \textbf{ID} & \textbf{Friendship Duration} & \textbf{Method of Friendship Initiation}                                             & \textbf{Platforms Used During Friendship}                                                                         \\ \hline
P1             & 6-7 years           & Minecraft server                                                & Discord, Kik, Minecraft, Phone, Skype, Snapchat, Steam, Teamspeak                                        \\ \hline
P2             & 1 year              & Mental health Discord server                                    & Discord, Genshin Impact, Texting, Voice and Video calling                                                \\ \hline
P3             & 5 years             & Mutual friend from Xbox Live                               & Discord, Team video games, Xbox Live chat                                                                \\ \hline
P4             & 2 years             & Sky: Children of the Light                                      & Discord, Sky: Children of Light, Texting                                                                  \\ \hline
P5             & 14 years            & Forum-based fan site                  & Discord, Email, Nintendo Switch games, Simpsons fan site                                                 \\ \hline
P6             & 1 year              & Writer and reader comment interaction on fanfiction site        & Archive of Our Own, Email, Tumblr, Whatsapp                                                              \\ \hline
P7             & 4.5 years           & VRChat                                                          & Discord, Drawing applications, Videos games, VRChat                                                      \\ \hline
P8             & 1.5 years           & Discord server for Dota 2                                       & Discord, Dota 2, Texting, Voice calls and texts                                                          \\ \hline
P9             & 6 years             & Drawing-centered Discord server                                                  & Discord, Drawing applications, League of Legends, VRChat                                                 \\ \hline
P10            & 3-4 years           & Interactions on Twitter and mutual Discord group chat           & Discord, Facebook messenger, Google drive, Physical mail, Twitter                                        \\ \hline
P11            & 12 years            & Blogger livestream event                                        & BlogTV, Phone calls, Skype, Texting, Whatsapp                                                            \\ \hline
P12            & 4 years             & Interaction on Instagram account for career activity              & Instagram, Phone calls, Texting, Video calls                                                             \\ \hline
P13            & 10 years            & Interaction on Livejournal blog for sports celebrity            & Facebook Messenger, Instagram, Livejournal, Twitter                                                      \\ \hline
P14            & 5 years             & VRChat                                                          & Discord, VRChat                                                                                          \\ \hline
P15            & 6 years             & Discord server for Minecraft Youtuber                            & Discord, Video games, Texting                                                                            \\ \hline
P16            & 5 years             & Discord server for Minecraft                                    & Discord, Minecraft, Physical mail, Texting, Voice calls                                                  \\ \hline
P17            & 3 years             & VRChat                                                          & Discord, Phone calls, Texting, VRChat, VR and video games                                                \\ \hline
P18            & 10 years            & Interactions on DeviantArt about fanart and a roleplaying forum & DeviantArt, Discord, Forum site, Texting                                                                 \\ \hline
P19            & 4 years             & Discord server for a specific character relationship pairing    & Discord, Physical mail                                                                                   \\ \hline
P20            & 7 years             & Mutual friend from a gaming group                               & Discord, Final Fantasy 14, Player Unknown Battle Ground, Teamspeak                                       \\ \hline
P21            & 6 years             & Writer and reader comment interaction on fanfiction site         & Archive of Our Own, Discord                                                                              \\ \hline
P22            & 7.5 years           & Interaction about fandom on Tumblr                              & Archive of Our Own, Discord, Skype, Tumblr                                                               \\ \hline
P23            & 12 years            & Tumblr roleplaying group                                        & AOL instant messenger, Discord, Jackbox Games, Phone calls, ProBoards forums, Texting, World of Warcraft \\ \hline
P24            & 10 months           & Discord server moderator group                               & Discord, Instagram, Overwatch, Whatsapp                                                                  \\ \hline
P25            & 3 years             & Writer and reader comment interaction on fanfiction site        & Archive of Our Own, Conversation starter website, Discord, Instagram, Netflix, Tumblr, Whatsapp \\ \hline  
\end{tabular}%
}
\caption{Details of Participants' Online Friendships}
\label{Tab:Meet}
\end{table}


\subsection{Interview Procedure}
\label{subsec:interview}
We conducted semi-structured interviews with all participants, starting with them telling us the story of their friendship with their close online friend. They then completed a visual elicitation activity, followed by a series of questions, both focused on the process of forming and maintaining their friendship. Each interview started with participants reconfirming their eligibility and consenting to participate in the study. Participants were asked to focus their answers on the online friend with whom they had the strongest connection for the interview. 

\subsubsection{Visual Elicitation}
During the visual elicitation activity \cite{johnson2002elicitation}, participants were asked to map out on a timeline (via Google Jamboard) the activities and technologies that played roles in their friendship formation and maintenance. The timeline encompassed the time from meeting their friend to the current time or the end of their friendship. The goal of this activity was to get participants to discuss not only the different activities and technologies used but also how and when they changed over the course of the participant's friendship. Framing the activity \textit{over time} highlights key points of transition during the participant's friendship.
\begin{figure}[h! tbp]
  \centering
  \fbox{\includegraphics[width=\linewidth]{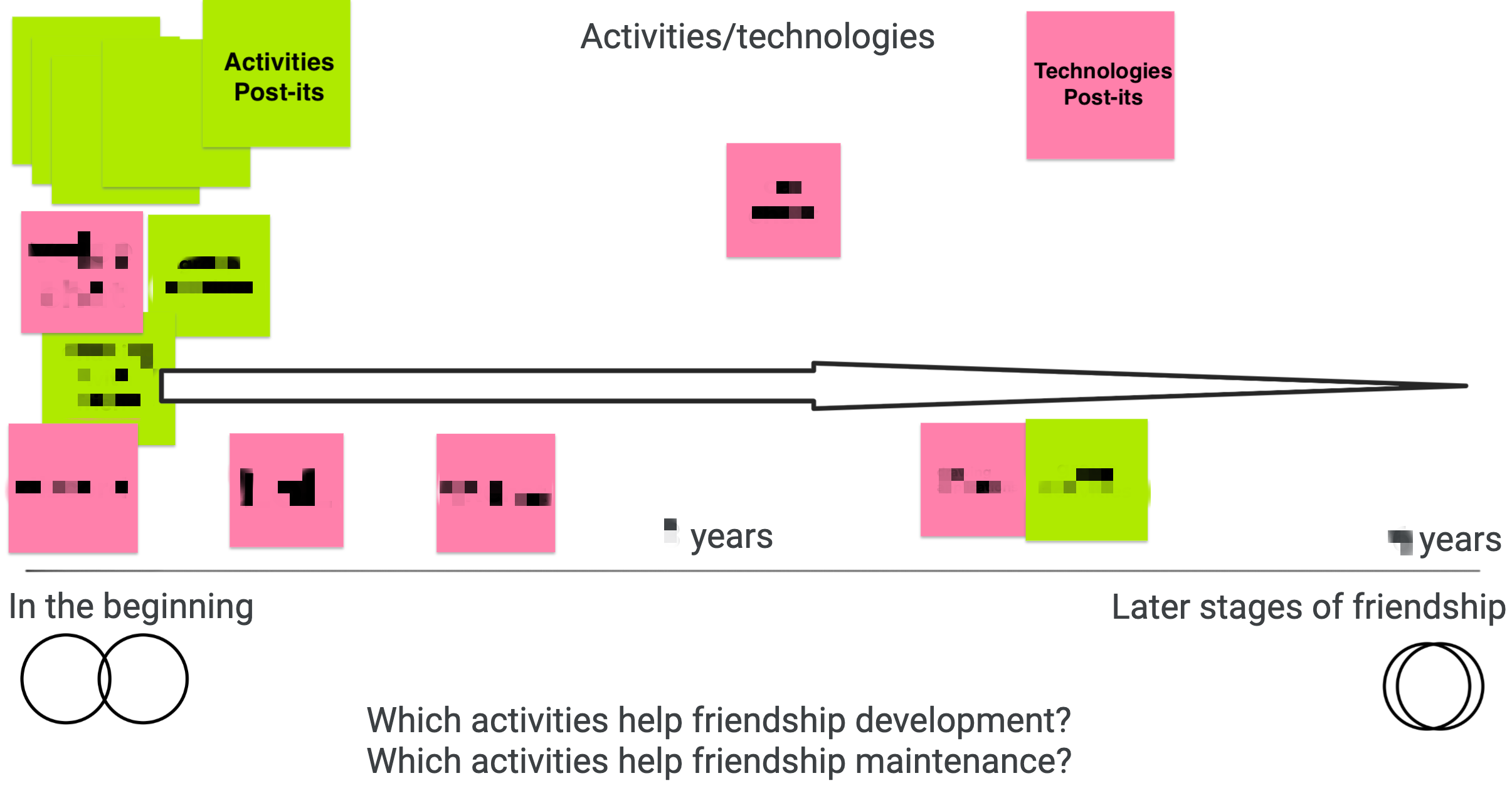}}
  \caption{An example of a visual elicitation. Participants mapped out and discussed which platforms and activities were involved throughout their online friendship.}
  \Description{Visual Elicitation timeline}
  \label{fig:visualelicitation}
\end{figure}

\subsubsection{Semi-Structured Interview}
Interviewers asked process-oriented questions about forming and maintaining participants' online friendship and the platforms they used to facilitate their connection. Questions focused on perceptions and goals of the friendship, as well as activities, behaviors, and struggles that occurred at transition points during the friendship.
\revision{For example, we began with ``Tell me the story of how you became close friends with this person you met online,'' which supported a process-oriented narrative; we followed up on described phases of the friendship with details such as ``What activities did you do, and what tools did you use for them?'' or ``What aspects of your environment (people or technology) affected your friendship at this point?'' which refer the {\small \texttt{transformation}}al process of AT and its sociotechnical {\small \texttt{elements}}.}
The questions were informed and adjusted based on the participant's story of their friendship and the visual elicitation activity. Doing so allowed interviewers to ground questions in participants' lived experiences and elucidate changes and tensions in participants' friendships. The interviewer would further listen for issues surrounding platform design and follow up on those struggles or request further information.

\subsection{Analysis}
\label{subsec:analysis}
Given our focus on the online friendship process and how these friendships manifest across varied sociotechnical contexts, an activity-grounded analysis of our interviews is best suited for identifying key {\small \texttt{tensions}} and points of consideration at different stages of friendship development. Charmaz's \cite{charmaz_constructing_2006} grounded theory supports use in conjunction with existing theoretical approaches to add context, and \citet{seaman_adopting_2008} outlines the mutually-complementary nature of grounded theory and activity theory: grounded theory\revision{'s focus on inductive analysis allows for insights to be constructed from the data in a bottom-up manner unrestricted by any imposed theories}, while activity theory \revision{is a flexible lens which} supports analyzing processes of change, particularly those entangled with the cultural environment of participants' lived experiences.

Our work adopts methodological guidelines for activity-grounded analysis, mainly \textit{theoretically-informed sampling} (Section~\ref{subsec:participants}) and \textit{process-oriented questioning} (Section~\ref{subsec:interview}) \cite{seaman_adopting_2008}. We specifically use constructivist grounded theory \cite{charmaz_constructing_2006} alongside activity theory \cite{leontev_activity_1978,engestrom_learning_2014} in this work. The first two authors open-coded and labeled common ideas across interviews to allow topics to emerge from the interview transcripts. Over the course of data collection, they discussed emerging topics until clear themes were established, placing an emphasis on {\small \texttt{elements}} and evolving {\small \texttt{tensions}} of the online friendship activity system.
All authors reviewed and discussed the final themes 
until a consensus was reached.
\revision{Examples of codes created at earlier steps of analysis included observations such as `tangible artifacts/affirmations reinforce presence' or `interests as fuel,' which coalesced into themes such as `qualities of experience that support persistence of friendship' that more broadly reflect links between friendship behaviors and sociotechnical dynamics.}

\section{Findings}
\label{sec:findings}

Our analysis aimed to understand the cultural-historical context of online friendship as an {\small \texttt{activity}} over time.
Our findings answer the research questions by discussing key aspects of online friendship development as {\small \texttt{elements}} of the sociotechnical activity system (See Fig. \ref{fig:activitydiagram}) \textbf{(RQ2)} and {\small \texttt{tensions}} within the system that reveal core characteristics of online friendship development \textbf{(RQ1)}.
We will use {\small\texttt{this notation}} to refer to elements and meta-features of the activity system in this work.

The first section (\ref{subsec:OF_Activity}) unpacks online friendship development as an {\small \texttt{activity}}, and the remaining sections (\ref{subsec:findngs_conditions}, \ref{subsec:findings_modality}, \ref{subsec:findings_onlinevsirl}) highlight key takeaways as {\small \texttt{tensions}} 
(originally defined as {\small \texttt{contradictions}} \cite{engestrom_learning_2014})
in the activity system, which drive dynamics of online friendship development. 
Each inner subsection will describe a critical {\small \texttt{tension}} impacting the success of online friendships.

Section~\ref{subsec:OF_Activity} introduces the {\small \texttt{elements}} of online friendship {\small \texttt{activity}} and illustrates how these interacting {\small \texttt{elements}} influence the development of online friendships over time.
Section~\ref{subsec:findngs_conditions} presents achieving a sense of \textbf{temporal constancy} in friendship interaction and the crossing of sociotechnical \textbf{turning points} as key challenges to overcome in developing close and long-lasting online friendships. 
Section \ref{subsec:findings_modality} presents detailed ways in which online friends interpreted and re-appropriated the sociotechnical affordances around them to co-create a sense of presence and connection in the virtual setting.
Lastly, Section \ref{subsec:findings_onlinevsirl} identifies the significant impact of real-life pressures on online friendships. Participants' discussions of stigma, norms, and examples of friction also clarify the context of public doubt towards online friendship.

\subsection{Online Friend Maintenance as an Activity}
\label{subsec:OF_Activity}

\revision{In this work we follow Engestr\"{o}m's extended AT model \cite{engestrom_learning_2014}, which articulates how {\small \texttt{tensions}} among {\small \texttt{elements}} and other structures in the activity system drive its {\small \texttt{transformation}} over time (See Section \ref{subsec:atintro}). Online friendship development is by definition an {\small \texttt{activity}} in AT due to being a social systemic process oriented towards a shared goal (increasing reciprocity or closeness between friends), evolving over time, and mediated by tools and cultural-historical factors \cite{engestrom_learning_2014,rubin_handbook_2009,amichai-hamburger_friendship_2013}.}

\begin{figure}[h! tbp]
  \centering
  \includegraphics[width=\linewidth]{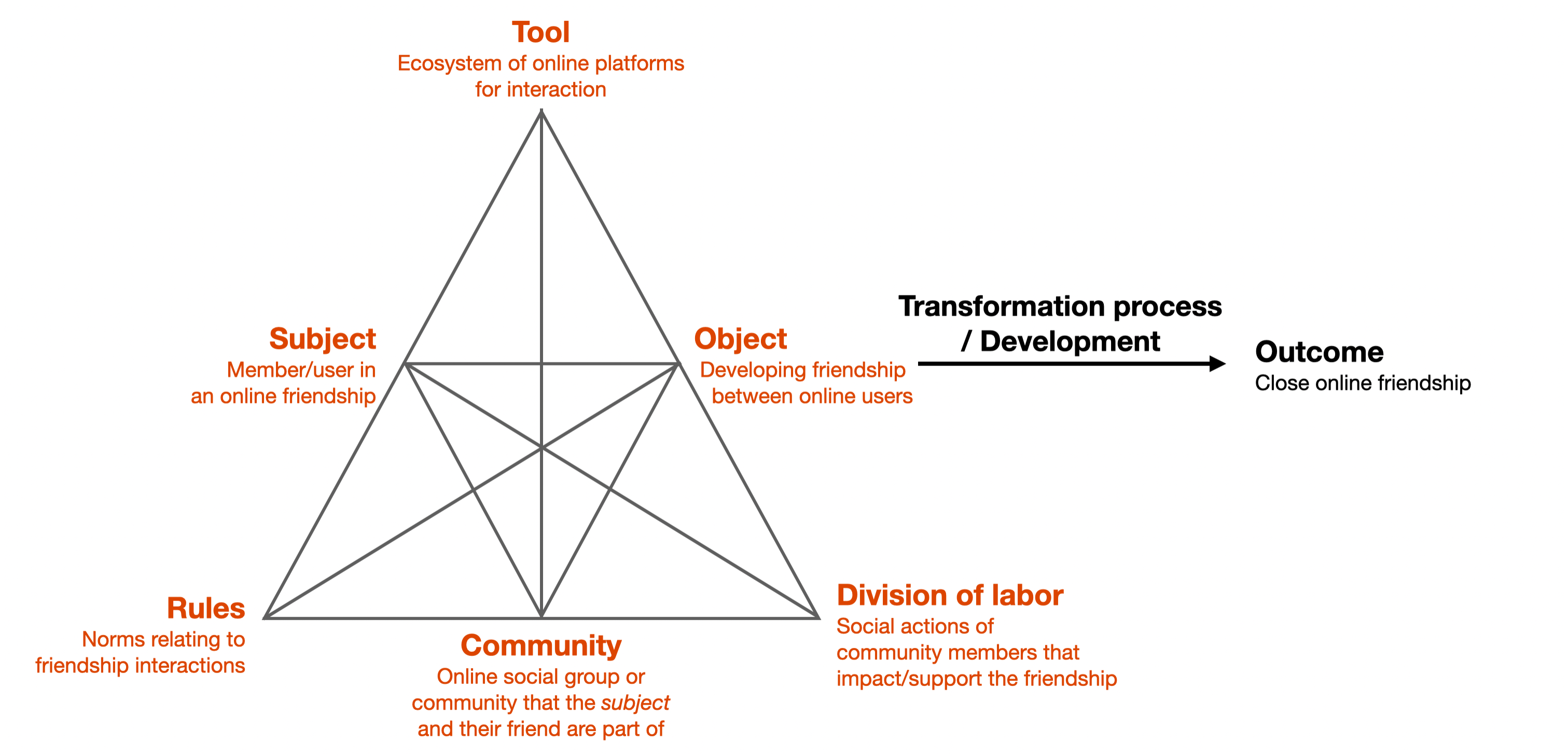}
  \caption{\textbf{The online friendship {\small \textbf{\texttt{activity}}} system.} The six \textcolor{RedOrange}{\small \textbf{\texttt{elements}}} are: {\small \texttt{subject}}, {\small \texttt{object}}, {\small \texttt{tool}}, {\small \texttt{rules}}, {\small \texttt{community}}, and {\small \texttt{division of labor}}. Other meta-features are the {\small \texttt{activity}} of online friendship development which contains all six {\small \texttt{elements}}, and the desired {\small \texttt{outcome}} as a successfully sustained and close online friendship.}
  \Description{Activity Theory Diagram}
  \label{fig:activitydiagram}
\end{figure}

Our analysis identified how each {\small \texttt{element}} of the activity system (Figure~\ref{fig:activitydiagram}) correlated to characteristics of the online friendship development process.
\begin{itemize}
    \item {\small \texttt{subject}}: a member of the online friendship being developed. Each participant and their described online friend are considered as individual {\small \texttt{subjects}}.
    \item {\small \texttt{object}}: the developing relationship between online friends.
    \item {\small \texttt{tool}}: the ecosystem of artifacts (both online and physical) afforded to online friends.
    \item {\small \texttt{rules}}: norms for interaction between online friends, including norms that originate internally or externally to the friendship.
    \item {\small \texttt{community}}: online social groups or communities that {\small \texttt{subjects}} are a part of
    \item {\small \texttt{division of labor}}: individual {\small \texttt{community}} members' influence on the friendship
\end{itemize}
The \textbf{process-oriented} nature of online friendship {\small \texttt{activity}} is shown through meta-features of the activity system (Figure~\ref{fig:activitydiagram}).
The desired {\small \texttt{outcome}} of friendship development (a close friendship) is the goal of the {\small \texttt{activity}} (the six-{\small \texttt{element}} system).
The {\small \texttt{transformation}} of the activity system characterizes the development towards the {\small \texttt{outcome}} over time (see Section~\ref{subsec:findngs_conditions} for highlights).


\begin{figure}[h! tbp]
  \centering
  \includegraphics[width=\linewidth]{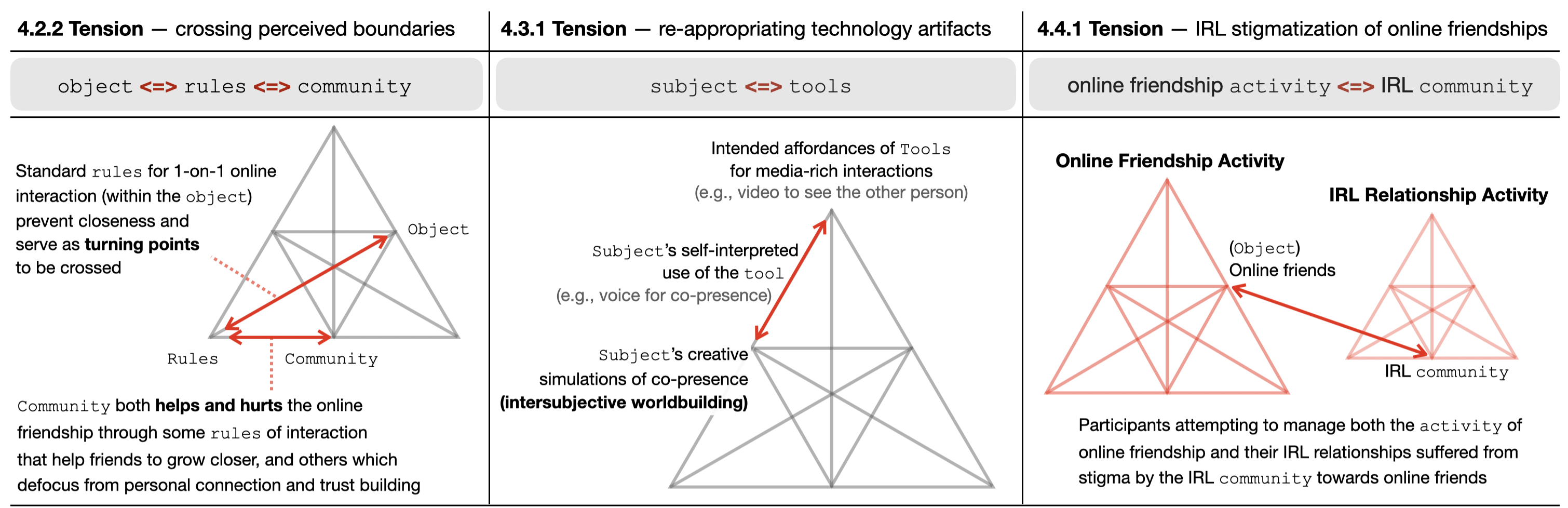}
  \caption{\textbf{Examples of {\footnotesize \texttt{tensions}} (marked in \textcolor{BrickRed}{red}) from Sections \ref{subsubsec:crossing_perceived_boundaries}, \ref{subsubsec:contextmatters}, and \ref{subsubsec:stigma}.} We present key takeaways as {\footnotesize \texttt{tensions}} to reflect how AT dynamics impact the development process for online friends.}
  \Description{Tensions Diagram}
  \label{fig:tensiondiagram}
\end{figure}

\newpage We illustrate the mapping of this activity system onto the online friendship process using one individual's story as an example: P22 ({\small \texttt{subject}}) was in a chat group ({\small \texttt{community}}) on a messaging platform ({\small \texttt{tool}}) with their friend. They described instances of awkwardness in their friendship ({\small \texttt{object}}) during which group members' continuation of their interaction ({\small \texttt{division of labor}}) normalized ({\small \texttt{rule}}) restarting the conversation with their friend. This contributed to the {\small \texttt{transformation}} towards their {\small \texttt{outcome}} by allowing the friends to sustain contact.
Here is only one case of how {\small \texttt{elements}} may interact, while the {\small \textbf{\texttt{tensions}}} (\autoref{fig:tensiondiagram}) described in the sections below depict more complex interactions in the {\small \texttt{activity}} of online friendship.

\subsection{Online Friendship Development: Reaching constancy through turning points}
\label{subsec:findngs_conditions}
Spanning experiences across various online platforms, all participants recalled reaching a point of feeling as if their online friend \textit{``had always been there''} as a part of their lives. As explained by P10:
\begin{quote}
    \textit{``You don't feel like anything massive ever happened. But suddenly like, a really important part of your life, you're like, `Wow that's nice. We're already close friends.'''}
\end{quote}
We call this a state of \textbf{temporal constancy} in the friendship {\small \texttt{outcome}}. Temporal constancy was a key aspect of closeness reached through sustaining friendship interactions over time, which involved crossing perceived boundaries to enable more personal interactions (\textbf{turning points}), and accumulating meaningful shared experiences.

The key {\small \texttt{tensions}} in this section highlight how norms ({\small \texttt{rules}}) of online friendship interaction are constructed or surpassed through various activity system {\small \texttt{elements}}: the online friend as a {\small \texttt{subject}}, the friendship {\small \texttt{object}}, the platform as a {\small \texttt{tool}}, online {\small \texttt{community}} members, and their {\small \texttt{division of labor}}.
Each {\small \texttt{tension}} (see Fig.~\ref{fig:tensiondiagram}) illustrates how friends construct their own norms of interaction (\ref{subsubsec:creatingtime}), move past certain norms of interaction (\ref{subsubsec:crossing_perceived_boundaries}), and select for situations which facilitate interaction norms most helpful to developing close online friendship (\ref{subsubsec:balancing_task_personal}).




\subsubsection{Creating time for consistent interaction}
\label{subsubsec:creatingtime}
The \textbf{temporal constancy} that brought a sense of stability and security to the friendship was cultivated through consistent involvement in each other's lives across a long span of time.
However, restrictions in time available for online friends ({\small \texttt{subjects}}) to develop their friendship ({\small \texttt{object}}) served as a {\small \texttt{tension}} which drove friends to develop routines of interaction ({\small \texttt{rules}}) which supported sustained interaction over time.

Such limitations on time included challenges such as residing in different time zones and physical obligations to family, work, or school.
P24 mentioned how considerable time zone differences sometimes prevented them from asking for advice when needed, 
and P23 described the difficulty of\textit{``just trying to stay friends and stay close''} during a new relationship and job.

Participants unanimously mentioned the solution as establishing routines ({\small \texttt{rules}}) of interaction built on their existing connection instead of trying to explicitly schedule consistent interactions with one another.
Routines provided a general awareness of friends' daily lives and schedules, which was necessary for coordinating the best times to spend together. P4 explained, \textit{``We message throughout the day, but he is an hour behind me so we have most of our evenings freed up to actually spend time with one another.''}
Time of day was also an important factor in developing routines of friendship interaction. Several participants preferred nighttime for more personal interactions:
P1, who was shy about speaking, explained that \textit{``At night, when everybody was calming down, it was easier for me to talk. The atmosphere changes; people get softer and less on-edge.''}
Others talked about the routine of continuing conversations at the start and the end of the day, and even those with major time zone differences had their own routine, as P24 explains: 
\begin{quote}
    \textit{``We kind of just keep talking on, the conversation topic changes, we don't really do like oh good morning or I'm going to bed, but sometimes I swear we're awake and go to bed at the same time... so \textbf{it kind of feels like he's just always there.}''}
\end{quote}


Developing a sense of personal closeness between friends created the necessary confidence to develop routine contact without fearing to initiate new or different things with an online friend.
P17 described how after becoming close with her friend, \textit{``We don't care if we don't have anything to talk about, we'll just send funny pictures or a smiley face. It's just, `Hey, I'm thinking about you.'''}

Progressive activities with history gave the strongest support for building more consistent routines of interaction among friends.
For example, P22 and their friend grew close by collaborating on a writing challenge where they edited and exchanged draft feedback for months, leading them to talk \textit{``every single day in private messages on Discord.''}
Some progressive activities also created \textit{tangible history} as a perceivable {\small \texttt{tool}} for friends to look back on, as illustrated by
P1's experience in Minecraft: \textit{``The sense that we're all contributing to progress as a whole. Building our own little village, seeing something that everybody works together to make.''}
P10 also mentioned the tangible value of archived collaborations with their online friend:
\begin{quote}
    \textit{``All of our shared history is [in those Google Drive drafts], in little bits and pieces that they've given me... It just makes me feel both really fond and like, \textbf{oh, actually, I think that we're both committed to having this last. It's like a paper trail.}''}
\end{quote}

Although achieving \textbf{temporal constancy} was challenging for online friends, our participants addressed this {\small \texttt{tension}} by developing routines ({\small \texttt{rules}}) which enabled them to sustain consistent connection. Shared, ongoing collaborative activities also produced tangible history as {\small \texttt{tools}} that supported the sense of \textbf{temporal constancy}, and allowed friends to strengthen the personal connection necessary for establishing regular routines of interaction.

\subsubsection{Crossing perceived boundaries to advance closeness}
\label{subsubsec:crossing_perceived_boundaries}
Establishing personal connection was crucial for achieving \textbf{temporal constancy} in online friendship. 
However, existing {\small \texttt{rules}} of interaction in both the 1-on-1 online friendship ({\small \texttt{object}}) and the social environment of online platforms ({\small \texttt{community}}) created the {\small \texttt{tension}} of perceived social boundaries that online friends had to cross to become closer.
Crossing these boundaries served as \textbf{turning points} for progression in the friendship. We also identified {\small \texttt{community}}, {\small \texttt{division of labor}}, and {\small \texttt{tool}} (platform) dynamics that affected the crossing of these boundaries.

Depth of personal connection enabled online friends to confide in and support each another, viewing each other as \textit{``the person I tell things to''} (P23).
Online friends supported our participants through difficult times, such as staying home to keep them company (P1, P22) or even arranging surprise events to cheer them up (P3).
However, participants described the need to cross boundaries such as transitioning into personal conversation topics or making the leap from texting to voice calling to develop closer friendships.
This transition was necessary, but difficult due to concerns about violating perceived boundaries. 

Making the first voice call was one major \textbf{turning point} for most participants. 
Voice calling was a crucial way to develop close online friendships (see Section~\ref{subsec:findings_modality}), and many participants considered initiating voice call in a 1-on-1 online relationship to be breaching a {\small \texttt{rule}}.
P7, who met his friend as a \textit{mute} player (users who do not speak) in VRChat, explained that having the mic off \textit{``was the norm. Talking would be the exception to the norm.''}
Other participants mentioned how \textit{``the point at which you hear someone's voice for the first time is a really important connection,''} (P10) emphasizing the emotional impact of this moment.

Certain factors helped to trigger the first voice call. Being part of a {\small \texttt{community}} which used voice call {\small \texttt{tools}}, such as a gaming group or Discord voice channel, made voice calling with a specific friend much easier. 
P22 elaborated how \textit{``joining the voice channel instead of having to call specific people''} was less stressful by softening social expectations.
P3 described how being able to go on the first individual voice call broke a barrier to engaging in these closer interactions more regularly: 
\begin{quote}
\textit{``One day there just wasn't anyone else there. And once you have that first duo interaction, it's like going on a first date with someone, and then it goes well, and you're more comfortable going on a second and a third.''}    
\end{quote}
Incidental reasons also triggered initial voice calls, such as providing emotional support during a difficult situation (P11) or keeping a bedridden friend company (P22).
Specific tasks, such as sharing project ideas or professional advice also spurred first calls.
P10 described how their first call as a physical therapy session turned into a deep conversation: \textit{``We were already friends. But it lasted an absurd amount of time, because we did the business stuff, and then we just continued talking.''}

Another essential \textbf{turning point} for online friendship development was transitioning from anonymous online acquaintances
to a close 1-on-1 friendship.
Here, membership in an online group had double-edged effects on online friends' ability to cross that line. This entailed online group members' involvement ({\small \texttt{division of labor}}) and interaction {\small \texttt{rules}} in the {\small \texttt{community}} setting.

Online friendships that developed in a `vacuum' instead of an active {\small \texttt{community}} setting struggled with intimacy and disclosure. 
P21 and her friend who \textit{``started anonymous with each other''} took \textit{``baby steps''} across several years due to \textit{``getting so used to that boundary.''} They could only self-disclose to each another incidentally, such as finding out about birthdays from a photo on the day of, or mentioning family during a scheduling conflict.

By comparison, online friends in a {\small \texttt{community}} or group reported having more natural opportunities to develop their personal connection without feeling like crossing an explicit boundary, especially since there was a \textit{sense of place} to social groups (P1, P10), 
providing serendipitous opportunities for participants to learn about a potential online friend and develop personal interactions.
P3 explained how \textit{``Observing how well people are able to get along with others, whether their friends or a stranger''} reflected upon the person and was important for developing trust towards an online friend. 
P22 also explained how {\small \texttt{division of labor}} from other group members enabled them to overcome hurdles in the closeness of their online friendship:
\begin{quote}
    \textit{``Online, you start cautiously, and you'll often hit a point where there is an actual conflict or disagreement. \textbf{It's much easier to not just fade out of each other's lives when the other people quickly take control of the conversation and keep it going.} That allows you step away and come back, and for there to not just be, if it's just the two of you, your last message is still whatever it was.''}
\end{quote}

Although {\small \texttt{communities}} supported advancing closeness, they also worked against it through {\small \texttt{community}}-embedded {\small \texttt{rules}} and platform ({\small \texttt{tool}}) limitations which hampered online friends' ability to foster personal connection.
Many participants established their friendships in tight-knit social groups that were difficult to access due to \textit{`cliques'} developing in online spaces. P9 described a series of \textit{``weird and bad''} social experiences during his journey to find the \textit{``tiny little Discord server''} where he met his closest online friend, and P14 reported a similar issue on VRChat:
\begin{quote}
    \textit{``The furries, adventurers, and artists went into their own groups. Now it's a little harder to make genuine friends in the space with people who are not already associating with anyone else at the moment. It's hard to find specific communities because they're already where they want to be.''}
\end{quote}
Participants also described how the overwhelming nature of high-anonymity, high-interactivity online social platforms such as larger Discord servers and VRChat venues promoted norms of insincere behavior. This made it difficult for online friends to trust or be vulnerable in front of one another.
P24, who moderated a large Discord server, described how:
\begin{quote}
    \textit{``Discord is pretty much a huge group chat with all these people, and you don't have to use your own face and name. It's easier for people to do and say things they probably wouldn't want people they know in-person to know they're doing.''}
\end{quote}
P20 experienced a similar lack of genuineness in public groups in VRChat:
\begin{quote}
    \textit{``You're role playing as a character. We were more guarded about our personal life; I can treat `VR people' with a degree of dissociation from a real person.''}
\end{quote}
\revision{Aside from lacking the trust necessary to build a friendship, participants also found it difficult to relate to others or have meaningful interactions in such large communities. P10 stated that \textit{``You don't exist in a society, and I mean that in the small sense you don't exist in a community.''}
P6 also shared that being in a large Discord server felt \textit{``like watching someone else's group chat. It feels like a friend group I'm not a part of.''} P22 explained this dynamic in detail:
\begin{quote}
    \textit{``\textbf{Being in a server helps a lot, but it also slows down friendship development,} because things only happen out of natural interactions. Somebody might like somebody else and want to be friends with them. But you're in a larger group of people, so you get to know them in a low pressure environment.''}
\end{quote}
This slower pace of friendship development was further exacerbated by \textit{``drama''} which nearly all participants reported experiencing at some point in these larger, highly-interactive communities. Social conflict or toxic behavior on these platforms heavily burdened friendships that were still developing on them. For example, P16 lost contact with their friend \textit{``because something tragic had happened in one of [online friend's] servers.''} Toxic community events led to negative mental associations towards using the platform, even to keep in contact with one's friend. P18 reported a similar experience when their friend was unfairly banned from the site they were both on.}

\revision{More broadly,} group- and interest-oriented dynamics in many online social spaces also detracted from focus on 1-on-1 connection. Several participants mentioned moving their online friendships off of the original platform as a result: P6 explained moving her online friend to Whatsapp as a \textit{``context switch''} from a fandom environment, and
P24 explained how \textit{``although you can privately chat on Discord, you're still constantly notified of everything else that's going on,''} demonstrating the extent of disturbance these spaces had on developing individual connections.


The {\small \texttt{tension}} in this section was illustrated through {\small \texttt{rules}} of 1-on-1 online interaction (in the {\small \texttt{object}}) and of the {\small \texttt{community}} which made it difficult for online friends to initiate closer relationships. 
We characterized these issues as \textbf{turning points} of (1) transitioning into voice-calling as a more personal {\small \texttt{tool}} for communication, and (2) transitioning from anonymous online {\small \texttt{community}} members into individual, personal relationships. 
Participants' complex discussions revealed how {\small \texttt{communities}} and {\small \texttt{tools}} (such as voice channels and {\small \texttt{division of labor}} in friend groups) provided opportunities and resources to overcome these \textbf{turning points}. However, online {\small \texttt{community}} and platform ({\small \texttt{tool}}) dynamics such as insincerity and group dominance could also work against the \texttt{outcome}.


\subsubsection{Personal interests and experiences as avenues to shared connection}
\label{subsubsec:balancing_task_personal}

Participant accounts revealed a {\small \texttt{tension}} between the {\small \texttt{subject}} and {\small \texttt{object}}: individuals' original interests for being on a platform, versus the need for social engagement to nurture online friendship.
Resolving this {\small \texttt{tension}} showed how certain elements of {\small \texttt{communities}} and {\small \texttt{tools}} fostered {\small \texttt{rules}} of interaction which supported the friendship {\small \texttt{outcome}}, namely by surpassing \textbf{turning points} in the friendship (\ref{subsubsec:crossing_perceived_boundaries}) and cultivating a sense of \textbf{temporal constancy} (\ref{subsubsec:creatingtime}).

Participants met their friends on platforms where the main goal was not necessarily personal connection, but rather a specific interest ({\small \texttt{subject}}), such as playing a game, moderating an online community, or being part of a fandom.
In contrast to the intuition of directly encouraging social engagement between online friends ({\small \texttt{object}}), participants reported that activities which linked back to their personal interests and motivations better supported the {\small \texttt{outcome}} by providing natural, less-superficial {\small \texttt{rules}} of interaction to accumulate shared experiences valuable to the friendship.

P4 explained how some MMOs attempted to support friendship through explicit `friendship tasks' such as finding another player and giving them a hug,
but battles and resource farming tasks with their friend, which were more in-line with a self-interest in the game, were more effective in drawing them and their friend closer. 
Instead of directly engaging with a friend, helping each other through a tangible experience gave them \textit{``something to talk about''} (P12) and \textit{``felt like a bigger deal than our usual interaction.''} (P5)

{\small \texttt{Subjects}}' personal interests played a strong role in fostering connection between friends.
Interest-centered interactions, whether in a topic, fandom, or hobby such as drawing, writing, or making, overwhelmingly served as \textit{fuel} for the relationship by propelling learning and self-disclosure.
P6 and P21 were fanfiction writers who became close online friends with one of their readers, and noted how their friends' in-depth interactions with the stories that they wrote allowed them to \textit{``get more of a sense of their [friends'] character.''}
Sharing interests in aesthetic topics also created pathways for self-disclosure between online friends.
P1 described, \textit{``I'll say, this song feels like this, or I associate this memory with this song. And then we start telling stories about each other's lives.''}
P6 also explained how fandom opened up vulnerable conversations: 
\begin{quote}
    \textit{``We'd be talking about, why do I like this character? Well, I like this character basically because of my daddy issues. And then we're just laughing about that. It's so intertwined, you end up talking about those personal motivations behind what you're writing and the content you're consuming.''}
\end{quote}
The connection of personal interests even spurred friends to develop mutual interest in one another, directly driving the friendship {\small \texttt{outcome}}.
For example, P17 and her friend were excited about and engaged in one another's hobbies even if it was not their personal interest:
\begin{quote}
    \textit{``He doesn't necessarily understand. Nor will he be as obsessed with making avatars as I am, and I wouldn't be as obsessed with this game that he absolutely loves. But the fact that we're sharing and talking passionately about something, I think that adds value.''} 
\end{quote}

Interests involving more interactive shared activities such as games ({\small \texttt{tool}}) also drove personal connection between friends, but certain mechanics were more effective than others.
Participants cited interest-based activities with natural opportunities for 1-on-1 interaction as most helpful for friendship, while an overwhelming focus on the task of interest detracted from it.
Creating time and space for individual connection was still important.
When recalling how their friendship began on VRChat, P17 said, \textit{``We did play games, but we liked exploring worlds a bit more, because we could talk one-on-one and not be focused on throwing a disc or playing cards with a bunch of people.''}
Participants who played League of Legends reported that its competitive nature and extreme focus on individual skills did not support personal bonding despite being a team-based game (P9,20).
In comparison, P20 described how battle royale games \textit{``focus on coordination rather than viewing people as above or under one's skill level, which help build implicit trust and cooperation in the relationship.''}

Though the {\small \texttt{outcome}} of online friendship was achieved through furthering engagement between friends (the {\small \texttt{object}}), we saw that starting from {\small \texttt{subject}}s' individual interests to motivate shared experiences and pathways to personal interaction effectively drove the friendship. 
However, {\small \texttt{rules}} encouraged by different activity mechanics exhibit mixed effects on bonding and demonstrate a need to balance between focus on individual interest and interpersonal connection.


\subsection{Co-Created Presence: Online friends' adaptation of sociotechnical affordances}
\label{subsec:findings_modality}
In an online friendship, available methods of fundamental communication are distributed across the {\small \texttt{tools}} available to {\small \texttt{subjects}} and vary along the dimensions of media modality (text, voice, video, or tangible object-sending) and synchronicity (asynchronous vs synchronous).
This section details online friends’ extremely creative interpretation and integration of sociotechnical affordances across different platforms ({\small \texttt{tools}}) in the act of defining and developing their friendship ({\small \texttt{object}}). 

The following subsections illustrate {\small \texttt{tensions}} between distinct affordances of available {\small \texttt{tools}}, the use of which online friends coordinated by considering the most meaningful aspects of their connection with one another ({\small \texttt{subject}} and {\small \texttt{object}}).
Specifically, these {\small \texttt{tensions}} highlight participants’ reappropriation of existing technologies based on their uniquely co-created sense of presence (\ref{subsubsec:contextmatters}), their coordination of synchronous and asynchronous interactions (\ref{subsubsec:reconcilingsynchronous}), and the permanence of effort expended in the friendship (\ref{subsubsec:effortful_interaction}).


\subsubsection{Re-contextualizing affordances through shared worldbuilding}
\label{subsubsec:contextmatters}
Although participants had to consider many factors when managing communication options, the way they adapted technologies and other features of their environment supported Walther’s claim \cite{walther_interpersonal_1992, walther_computer-mediated_1996} (Section~\ref{subsec:internetfriends}) that online interactions may not align with traditional face-to-face interaction, and actually surpass it.
Online friends' appropriations of technology reflected a form of \textbf{intersubjective worldbuilding}, where pairs of friends projected co-created meaning onto the affordances in their environment.
This instantiated the {\small \texttt{tension}} between {\small \texttt{subject}} and {\small \texttt{tool}}: online friends ({\small \texttt{subject}}) adapted sociotechnical affordances ({\small \texttt{tool}}) in ways that counter traditional concepts of use, reflecting very personalized interpretations of social presence.

Countering implications of media richness theory \cite{daft_organizational_1986}, our participants overwhelmingly preferred voice-based interaction over video calling for maintaining online friendships.
Voice calling enabled more flexibility for online friends to simulate natural ways of interacting which enhanced their sense of social presence.
Nearly all participants mentioned the importance of being able to hear someone's voice showcased here by P24: \textit{``You might know someone. But once you speak to them verbally, they feel like a real person.''}
Participants described how voice made it easier to convey and notice more complex emotional states, open up about personal topics, and understand their online friends' personality and sense of humor. P25 noted that voice was much better for conversations where \textit{``it's more about how you say it''} than about the surface-level content.

Online friends used voice calling to simulate co-presence by projecting shared mental models of the experience.
Voice communicated enough personal detail and the lack of visual input created room for participants to simulate physical presence with their friend. Several participants described the habit of voice chatting with their friend in the evenings while lying in bed as \textit{``kind of like a sleepover''} (P4). P8 described how when she placed her phone somewhere nearby, \textit{``If you don't look to the side, you could imagine that they're just laying right there next to you.''}

Online friends also developed proxy interactions over voice chat by anthropomorphizing the {\small \texttt{tool}}.
P22 described, \textit{``I might cook as we're talking on voice chat, then I'll be like, I'm bringing you to the kitchen!''}
They and their online friend also simulated physical contact through technology, in this case via a laptop:
\begin{quote}
    \textit{``Discord didn't always have noise crisping, so you could hear if the other person was typing and all those basic noises. We started doing this thing where we would tap on the shell of our laptop as a way of saying, `I'm here. I'm not saying anything,... but I'm right here and I care about you.' \\Both of us like to pet or hug people when they're distressed and we didn't want to just say `I'm hugging you,' because that defeats the purpose of a hug. So then \textbf{`Oh, I'm petting you' became `Oh, I'm petting the laptop which has you in it.'}''}
\end{quote}
Voice calling also allowed online friends to find comfort in spending time together: \textit{``to just sit in silence and just do our own thing, and every once in awhile say something to each other''} (P21). This additionally helped stimulate consistent interaction since it reduced digital fatigue and allowed participants to do something in addition to talking with their friend (Section~\ref{subsubsec:balancing_task_personal}).

Although video was a richer communication medium than voice-only, the added visuals were not preferred by most participants.
Video was only used on occasion to \textit{``show each other what you are doing''} (P8) or if friends \textit{``hadn't seen each other's faces for a particularly long time.''} (P12) Generally, participants did not find their online friends' appearance to be an important part of their emotional connection. As P8 explained, \textit{``I don't really care if the video is on or off. When we talk, the most important thing is being able to hear someone's voice and intonation. That's how I feel more connected.''}
Multiple participants mentioned how the social expectations ({\small \texttt{rules}}) associated with video calls, such as having to stay in front of the camera during the call, actually limited the natural activities they could do while talking to their friend. P22 explained, \textit{``Because of the fixed perspective, and you have to make eye contact in three dimensions through a screen when there's a person in front of you there's always this element of distress.''}

The favorability of voice and less immersive {\small \texttt{tools}} for maintaining and developing close personal connections with online friends also extended to VRChat-originated friendships.
Although VRChat interaction was rich and highly-immersive, participants' accounts reflected that it did not sustain long-term friendship development (the {\small \texttt{outcome}}).
None of our participants who reported meeting their close online friend on VRChat maintained the friendship through that platform due to its session-locked nature.
P14 also reported how getting accustomed to the intensity of social cues provided by VRChat caused him to lose contact with his close online friend, as it was difficult for him to adapt to communicating with his friend on other platforms. He explained,
\textit{``You get more out of VR than you do in Discord. You hear a voice, you see facial reactions, you see bodily reactions. You can still do it in Discord, but there's an element that obviously is just missing.''}

Participant accounts demonstrate how the shared context between online friends ({\small \texttt{subject}} and {\small \texttt{object}}) was much more important to their selection of communication {\small \texttt{tools}} than the surface-level richness of the {\small \texttt{tool}} itself. Online friends valued flexible forms of communication which could be sustained over a longer time. The strength of their personal connection was able to facilitate co-created presence even in the absence of structure, by combining features of available {\small \texttt{tools}} with altered interpretations of their physical environments. 


\subsubsection{Coordinating synchronicity of interaction}
\label{subsubsec:reconcilingsynchronous}
This subsection illustrates a {\small \texttt{tension}} within synchronous and asynchronous communication as mutually-exclusive affordances of {\small \texttt{tools}}. Though synchronous communication was desirable and allowed deeper engagement, online friends also learned to leverage unique benefits of asynchronous communication in order to sustain a long-term friendship {\small \texttt{outcome}}.

Participants highly valued synchronous communication (whether through voice, text, or other means) for its ability to make people feel emotionally closer and more present with one another.
P25 explained how synchronous conversations allowed for more personal engagement compared to asynchronous texting:
\begin{quote}
    \textit{``When we take days to get back to each other... It's not quite as spontaneous. \textbf{In a conversation going back and forth, a humorous moment just lands differently than it would with something that I had written a week ago.}''}
\end{quote}

However, as shown in Section~\ref{subsubsec:creatingtime}, participants could not engage in synchronous sessions of communication all the time, and asynchronous methods were still necessary not only to coordinate future synchronous sessions but also to maintain a persisting sense of closeness between friends.
In addition to providing online friends with more chances to contact each other, asynchronous media was valued for its ability to embody forms of expression and communication that persisted over time. P7 described exchanging drawings with his friend on an older and beloved feature of VRChat:
\begin{quote}
    \textit{``\textbf{The Presentation room} was very special, because you could join later - half an hour later, hours later, and you could still see all of their drawings, even though they were made long before you joined. The Presentation room was so beloved because VRChat didn't have a persistent message system.''}
\end{quote}
Because asynchronous communication {\small \texttt{tools}} served as “places” to leave messages, they eventually became likened to physical spaces ({\small \texttt{rule}}).
P4 described their close-friends-only Discord server as:
\begin{quote}
    \textit{``\textbf{A house server} where every channel is a different room. Each one of us has our own room, where we just talk about things that are on our mind. And then everybody has a place where you can reply to the stuff that they post inside of their channel.''}
\end{quote}

All participants strongly preferred sticking to a single {\small \texttt{tool}} for consistent, asynchronous communication with their online friend, which further reinforced the sense of having a concrete place of residence. P21 spoke of their shared private channel as: \textit{``A space where we can just be ourselves. No one else is watching, so we can do whatever.''}
This sense of place when anchored to an asynchronous platform strengthened the friendship connection ({\small \texttt{outcome}}) by allowing online friends to feel more grounded in a virtual world that may sometimes feel impermanent (see Section~\ref{subsubsec:stigma}).

Although the immediate closeness of synchronous interaction may often be foregrounded in the pursuit of deeper connection, its {\small \texttt{tension}} with asynchronous communication showed that such {\small \texttt{tools}} also provided unique and necessary value for online friendships by supporting
\textit{time-stable} social expression and providing \textit{a sense of place} crucial for `virtual' friendships.

\subsubsection{Preservation of effortful interaction}
\label{subsubsec:effortful_interaction}
Maintaining a close friendship only through technology-mediated means required a considerable amount of {\small \texttt{subjects}}' effort in the form of invested thoughts and actions.
\revision{Although participants did not explicitly describe their friendship as `effortful,' they willingly put in time and attention to sustain these valued connections.}
Participants' stories about the burden and benefits of such effort showed a {\small \texttt{tension}} within the {\small \texttt{subject}}, and how effortful interactions mediated through artifacts ({\small \texttt{tools}}) contributed to the {\small \texttt{outcome}} by lending a sense of permanence to online friends' virtual connection.

Relating to the ease and immediacy of interaction highlighted in Sections~\ref{subsubsec:contextmatters} and ~\ref{subsubsec:reconcilingsynchronous}, participants also reported the additional effort required to sustain certain forms of interaction: such as maintaining the mental context of a text-based conversation compared to being more present through a synchronous voice call. 
As P18 described, \textit{``[With a call] you can have a longer conversation in a shorter amount of time instead of typing out a whole wall of text.''}

However, modes of communication that inherently required more effort from the {\small \texttt{subject}} also manifested forms of social effort that enhanced the level of bonding in online friendship.
Text histories allowed for reflection on past interactions which, in the case of P10, reminded them that their friend \textit{``did all of these really sweet things for me''} and was \textit{``very nostalgic''}. The effort that unwittingly went into preserving these interactions allowed them to look back and more clearly appreciate the long-term value of the friendship.

Several participants also engaged in modes of communication with their online friend that were intentionally effortful and tangible, such as sending small gifts to one another and writing letters.
These temporally- and physically-persistent tangible affirmations acted as a \textit{``physical representation of the fact that, this person and I do care about each other,''} (P16) and \textit{``is a little piece of them... It's a connection that you are holding on to''} (P10). These forms of communication held immense value for participants' online friendships.
P19 told a story of feeling touched after unexpectedly seeing their gifts in a photo of their online friend's desk, which their friend said was reserved for \textit{``stuff that they treasure a lot and really value.''} 

Although {\small \texttt{subjects}}' effort in online friendships involved both compulsory effort (e.g., maintaining basic communication over text or preparing digital media), and more self-motivated effort (e.g., curating and exchanging gifts and letters), the artifact-mediated nature ({\small \texttt{tool}}) of these effortful actions allowed participants to generate permanent anchors of connection in the online friendship, strengthening the {\small \texttt{outcome}}.


\subsection{Online Friendship vs. the IRL Social Sphere}
\label{subsec:findings_onlinevsirl}
Engestrom’s model of activity theory \cite{engestrom_learning_2014} defines the highest level of {\small \texttt{contradiction}} (structural tension in the activity system) as a {\small \texttt{quaternary contradiction}} between the main activity system itself (the online friendship development) and other neighboring activities. 
Participants observed this type of fundamental {\small \texttt{tension}} occurring between their {\small \texttt{activity}} of online friendship development and the {\small \texttt{activity}} of managing IRL social obligations which were unavoidably a part of their lives. We abbreviate this neighboring area of activity as the `IRL social sphere.'

Each of the following subsections will describe this {\small \texttt{tension}} between online friendships and the IRL social sphere at different levels ---
\ref{subsubsec:stigma} and \ref{subsubsec:norms_of_interaction} depict societal and normative mechanisms of this tension, while \ref{subsubsec:frictionbetween} depicts grounded instances of conflict between online friendship development and IRL social obligations.
\ref{subsubsec:stigma} demonstrates the negative impact of stigmatization imposed by participants' ``IRL {\small \texttt{community}}'', but also illustrates participants' strong response of contrasting beliefs about online friendships.
\ref{subsubsec:norms_of_interaction} highlights differences between online and IRL social spheres in expected {\small \texttt{rules}} of interaction for maintaining friendship.
\ref{subsubsec:frictionbetween} then presents actual instances of conflict between online friendship development and the management of IRL roles and relationships.


\subsubsection{Stigmatization of online friendship}
\label{subsubsec:stigma}
One aspect of the {\small \texttt{tension}} between online friendship development and the IRL social sphere took the form of stigma towards online friendship from the ``IRL {\small \texttt{community}}'' (see \autoref{fig:tensiondiagram}, far right; this is separate from the {\small \texttt{community}} in \autoref{fig:activitydiagram}). Participants were, in response, even more driven to cherish and defend their online friendships. 
They not only shared vivid experiences of invalidation by people in their daily lives, but also explained how online friendships played a level of importance in their lives similar to IRL relationships.

P6 talked about how people in her IRL sphere looked down on her online friendship, with coworkers making comments like \textit{```Oh, that person's off with their `internet' friends. Yeah, they're weird,'''} and her husband saying things like \textit{```Why are you talking to this lady that you met because of your porn that you write online? That's not normal.'''}
P6 further stated, \textit{``There's this implication that if you have to make friends online, there must be something so weird about you that you can't make friends in real life. That's not even true.''}
P10 also elaborated P6's point about how the stigma was driven by a lack of understanding for the depth and complexity of online friendships:
\begin{quote}
    \textit{``\textbf{They're imagining more surface interactions}, anonymous user names, and people who don't really know each other, and they’re not imagining - Oh, yeah, I talk to them. I know this person's roommate, their wife, the name of the little kid that this person's wife nannies for. You know, all of these really personal things. They're not envisioning, you know, phone calls while we both do our laundry.''}
\end{quote}
Apart from P6, many other participants also felt anxiety and shame when online friendship came up in their ``IRL {\small \texttt{communities}}''. P13 recalled, \textit{``I remember being very judged by my family''} when sharing her decision to go on a trip with her online friend.

Despite this pressure, participants emphasized their unchanging perception of the online friendship ({\small \texttt{object}}): that the quality and priority of their online friendships were no less than with close others IRL. 
For example, P21 described how significant of a role her online friend played in her life: \textit{``She's one of the best things that's ever happened to me in my life. It's like having one pillar of your house that's always been there. If that pillar disappears, then that part of the house just falls away.''} P11 and P25 also described that their close online friend had, respectively, the same level of importance as their family members and childhood friend. 
In fact, the inherent fragility of online connection also drove participants to treasure their friendships even more out of a sense of urgency.
P21 spoke about how a sense of fear compelled her to actively connect to her online friend:
\begin{quote}
    \textit{``It was just, I really like talking to this person, and I also had a fear that, what if I can never see her again? What if the website goes down? I really like interacting with this person. I wanted a closer bond, so they don't just disappear into the Internet void.''}
\end{quote}

Online friends also pointed out how their regard for each other transcended online or IRL contexts.
P23 stated, \textit{``If she calls me, everything else gets dropped because I need to talk to [her] and make her feel better right now.''} Others provided IRL support for their friends, such as sending money, purchasing resources, and helping with remote physical tasks. 
Participants also emphasized valuing their online friends as unique personalities, which they did not see any differently than how they would someone IRL. 
P15 explained:
\begin{quote}
    \textit{``IRL, some personalities just click. I think it's the same way online, it's just a different medium for making friendships. \textbf{That doesn't mean the friendships are any less.} I feel like I’d receive the same feelings if I pulled that same prank on her in real life.''}
\end{quote}
Others made similar comments about the friendship development process, such as P17's account of getting to know her online friend through VRChat world-hopping:
\textit{``Most friends I’ve met are circumstantial, both IRL and online. We just hung out and talked - it’s like going on a hike with a friend. But only there are robots or stars instead of, you know.''}

Online friendships were deeply treasured despite the stigma they experienced from both close others and the wider public, and participants took care to explain how core aspects such as priority, loyalty, and feelings of friendship were no lesser in comparison to the IRL context.

\subsubsection{Norms of interaction}
\label{subsubsec:norms_of_interaction}
In comparing their experiences managing online vs. IRL friendships, participants also identified a {\small \texttt{tension}} in social {\small \texttt{rules}} regarding friendship interaction between the IRL and online context.
These differences in norms were sensitive to whether the friendship began in an online or IRL environment.
Even for the same person, their accustomed norms of interaction for online friendships were different from the ones they had for friendships formed in-person.
P11 recalled how compared with their online friend, they struggled to keep in contact with their in-person friends during COVID:
    \textit{``With [online friend], it's always just text and call, that's just how we operate.}
    \textit{\textbf{But it's weird to text some of my friends and the end goal not be a meetup.}''}
They described how this difference came from how the relationship began:
\begin{quote}
    \textit{``I’m so used to communicating with them with body language, and being able to see how they’re reacting to me. With [online friend], that's not a “need to know” thing when I'm talking with her. \textbf{It's more difficult to start off in person then have to go off of it, rather than learning to navigate somebody without it entirely.}''}
\end{quote}
This experience was echoed by others, such as P10 who maintained a close online relationship with their online friend but could not do the same with an IRL friend of six years, because of \textit{``the activities we had been doing together - we get used to climbing together and other things.''} 
P22 also recalled enjoying hanging out over long voice calls with their close online friend, but somehow could not do the same with their best friend since middle school: 
\begin{quote}
    \textit{``\textbf{We just never learned how to exist next to each other without being in person.} We never learned how to talk to each other over text because it was like, Well, we're gonna see each other at school. So from very early on, we just were not online friends.''}
\end{quote}

Participant comparisons of the online friendship development context with IRL situations revealed how the in-person accessibility and preordained concrete structure of IRL environments (as {\small \texttt{tools}} for socializing) tended to reduce intrinsic motivations to reach out asynchronously or through online means.
This dynamic contributed to a perceived {\small \texttt{rule}} in IRL social contexts of intimate online communication as counter-normative, fanning the stigma around online friendship interactions.
P11's earlier remark about the \textit{`weirdness'} of texting with her IRL friends for the sake of talking reflected a sentiment that was shared by other participants.
P19 perceived a \textit{``filter''} between herself and a prior close IRL friend when considering online interaction, \textit{``because we met from being in the same class. The main reason we hung out was to do work together and then we would chat once we were done.''} The concrete structure provided by that setting served as an IRL {\small \texttt{tool}} which grounded their friendship to that context.
P13 also mentioned not wanting to \textit{``bother family or [IRL] friends [by talking remotely], since I'm more likely to catch up with them later.''}
Although people IRL may have been physically closer to one another, their perceived social affordances for maintaining friendship were often restricted to those contexts and contributed to uncomfortable or negative perceptions of online-grounded intimacy.

These divergent {\small \texttt{rules}} of online vs. IRL personal interaction stemmed from the specific virtual or physical setting ({\small \texttt{tool}}) in which the relationship formed, constituting a system-level {\small \texttt{tension}} between the {\small \texttt{activity}} of online vs. IRL relationship development.
The sense of discomfort around translating acts of closeness meant for an IRL context into a virtual online context which we observed here may shed light on the stigma of perceiving intimate online-only friendships as abnormal or weird.

\subsubsection{Friction between online and IRL relational spheres}
\label{subsubsec:frictionbetween}

The significant {\small \texttt{tension}} between online friendship and the IRL sphere also occurred in the form of conflicting demands for {\small \texttt{subjects}}' personal resources, directly impacting participants' ability to engage in the {\small \texttt{activity}} of online friendship development.
Online friends experienced direct conflict with IRL social obligations that took precedence over friendship-maintaining processes, impacting communication and time spent together. This was the most significant source of high-risk conflict cited by participants, due to it invalidating the perceived importance of the online friendship.
Participants described how they were expected to, in practice, separate and demote their online social lives in comparison to their IRL physical connections. In light of these conflicting obligations, they also described their struggles and attempts to bridge this priority gap.

Participants reported how people in their lives expressed the view that it was not acceptable to give online interactions equal or greater priority to anything IRL, even if that was what resonated with them.
P3 explained, \textit{``I feel like a lot of people view online friendships as secondary to real ones. Whereas, I view all of my friendships on an equal level. As in, I would put the same effort as I do into his friendship as I would with somebody IRL.''}
P6 described the difficulty in legitimizing her own choice to spend time with her online friend by describing her husband's reaction to it: \textit{``He would not like that. It would be like, you're taking time away from me to talk to somebody online?''} She also described feeling a similar insecurity when her close online friend suddenly stopped talking to her:
\begin{quote}
    \textit{``I didn't know if she was losing interest, or what was going on. If she hadn't rekindled it, I would've been like. Oh, that was a great couple months of having a really close friend, but guess she got bored of me... it did hurt a little bit. Part of me was thinking, \textbf{I shouldn't care this much about this. But I do.}''}
\end{quote}


Compounded with real-life obligations, this pressure led many online friends to struggle with the {\small \texttt{activity}} of maintaining their friendship.
P23 described a difficult time in her online friendship when she felt the need to focus on developing IRL relationships: 
\begin{quote}
    \textit{``The shift in our dynamic caused some conflict, where [online friend's name] felt kind of left out... Because my husband was the first serious relationship in my life I probably also talked about it too much and made her feel like I didn't care about her anymore.''}
\end{quote}
P16 also recalled times when his online friendship \textit{``sort of waned,''} recalling \textit{``the waiting part, where things got very busy, and we just wouldn't really contact each other for a span of time.''}
 
Most participants talked about different IRL social obligations which made maintaining online friendships difficult. For P6's online friend it was family responsibilities as a mother of several children, for P17 it was getting a job with a different work schedule, and for P11 it was dealing with a college environment.
It was also easy for IRL social obligations which suddenly came up to be prioritized over time that online friends originally planned to spend together. One example P3 gave: \textit{``Say your mom comes home and she's like, `I need you to help put the groceries away,' then you have to be like, Oh, shoot, I gotta go.''}

In light of the struggle to prioritize online friendship, participants emphasized the importance of being mindful of the fragility of online presence and the {\small \texttt{activity}} of earnestly communicating necessary absences to online friends.
As seen from above, lack of communication around absences led to conflict or waning in the friendship. 
P25 explained the importance of updating her online friend during times of absence:
\begin{quote}
    \textit{``I don't like it when people feel like I've just been ghosting them for no reason. So I always prefer to say... it's not because I'm not interested, it's because I'm away. Because I... didn't want her to feel that I had dropped that interest.''}
\end{quote}
Other participants also mentioned how expressing sincerity and respect towards their online friends enabled them to stay close to one another despite interruptions or obligations from real-life.
Some appreciated that their online friends were \textit{``apologetic''} about replying late (P7), while others checked in with each other consistently (P11, P24).

Only two of our participants described being able to avoid conflict with their IRL relationships when maintaining online friendships. P17 and P23's close others IRL also had online friendships, and were accepting of integrating their online and IRL relationship {\small \texttt{activity}}. P17 planned a road trip with her husband to visit both of their online friends, and P23 described how \textit{``My husband's online friend group and mine combined in the last few years. We get together and play Jackbox games around once a month, where all of our friends come and we're all on a Discord call together.''}

Participants' varying experiences in coordinating their online friendships and IRL obligations showed how although these two largely separate {\small \texttt{activity}} systems experienced friction in competing for online friends' attention, 
there were exceptions to the stigma of online friendship as inferior and separate to IRL interaction.
In line with Engestr\"{o}m's concept of {\small \texttt{contradiction}} in activity systems as a key to their \textit{expansion} \cite{engestrom_learning_2014}, overcoming the beliefs (\ref{subsubsec:stigma}) and norms (\ref{subsubsec:norms_of_interaction}) which separate online friendship from the IRL social sphere could help \textit{expand} online friendship to be an activity more socially acceptable and integrated with real-life.

\section{Discussion}
In the following sections, we provide an overview of the principal findings before highlighting key extensions to the literature on online friendship, online community platforms, and computer-mediated communication.
For each, we follow with recommendations for future research and design of online social platforms and relationship maintenance.
\subsection{Principal Findings}
\label{subsec:principal_findings}
Our platform-agnostic investigation of online friendship experiences revealed both identifying characteristics of online friendship development (\textbf{RQ1}), as well as how it is impacted by the surrounding sociotechnical ecosystem (\textbf{RQ2}). An activity-grounded perspective allowed us to answer these research questions by identifying {\small\texttt{tension}}ed elements and meta-features of the activity system which drove the {\small \texttt{transformation}} and development of online friendship. 

A developmental focus on the activity system's {\small \texttt{transformation}} led us to identify \textbf{temporal constancy}, a permanent sense of online friends' presence, as an important characteristic (\textbf{RQ1}) of successful online friendships. In working through the {\small\texttt{tension}}s occluding this goal, online friends had to sustain consistent interaction (\ref{subsubsec:creatingtime}) and cross \textbf{turning points} (\ref{subsubsec:crossing_perceived_boundaries}) in both technology use and interactional norms.
Another characteristic was online friends' co-development of interaction norms and routines based on their unique shared sense of presence. Some chose not to acknowledge time-zone differences (\ref{subsubsec:creatingtime}), while others treated their communication devices as the other person (\ref{subsubsec:contextmatters}) or adapted online platforms as a place to stay (\ref{subsubsec:reconcilingsynchronous}).
The sense of urgency and earnestness in maintaining online friendships was also distinctly characterized by opposing pressures from the IRL social sphere (\ref{subsec:findings_onlinevsirl}).

The activity system framing also allowed us to identify influences from the sociotechnical ecosystem (\textbf{RQ2}), both as {\small \texttt{elements}} within the activity system and {\small \texttt{tension}}s external to the system (\ref{subsec:findings_onlinevsirl}).
We identified how online {\small \texttt{community}} membership could facilitate aspects of friendship development, but also propagate norms which discouraged interpersonal closeness (\ref{subsubsec:crossing_perceived_boundaries}).
{\small \texttt{Tools}} in the form of both online platform mechanics and affordances in the real-world environment (e.g., time of day and physical device setup) also impacted how online friends developed their relationship (\ref{subsec:findngs_conditions}, \ref{subsec:findings_modality}).
Lastly, the external {\small \texttt{tension}} exerted by the IRL social sphere as a {\small \texttt{quaternary contradiction}} (\revision{a {\small \texttt{tension}} resulting from conflict with neighboring {\small \texttt{activities}}, see} \ref{subsec:findings_onlinevsirl} \revision{and Fig. \ref{fig:tensiondiagram}[right]}) to the {\small \texttt{activity}} system was another key influence of the sociotechnical ecosystem that particularly distinguished online-only friendships from other contexts of friendship.

\subsection{\revision{Deconstructing} Stigma: Contextual Imprinting in Relationship Development}
\label{subsec:contextual_imprinting}

Participants' differing interaction norms for online and IRL friendships stemmed from the environment in which the relationship was formed (see \ref{subsubsec:norms_of_interaction}).
This difference in norms can extend our understanding of online friendship and its stigma by explaining mixed findings in prior work regarding the inherent quality of online friendships.

Our results concerning the sociotechnical ecosystem of online friends allow us to hypothesize about prior findings on online friends and their ``inferior'' quality. Studies observing inferior quality in online vs. IRL friendships were conducted in settings that confounded their context, such as sampling both from the same social networking site \cite{antheunis_quality_2012} or by studying the transition from IRL to online friendship \cite{scott_social_2022}. While valuable research, we posit that these studies may lead to false comparison between IRL and online friends, putting IRL friends at an advantage. Another study by Chan et al. aligns with our conjecture -- they explored separate reports of online and IRL friendships and observed similar quality between them \cite{chan_comparison_2004}.
Our findings in \ref{subsubsec:norms_of_interaction} show how participants' standards for friendship interaction were connected to the context of friendship formation (e.g., on an online platform vs. a college campus), which could explain these conflicting results in past work.
These results also align with the findings in \citet{lee_friendship_2024}'s post-COVID work on online friendship formation in graduate students, which found differences across context of friendship formation.
Based on participants’ experiences, the friction between real-life and online friendship related more to ill-understood distinctions in social context and less to the concept of an inherent deficiency in online friendship (the source of stigma) \cite{chayka_lets_2015,amichai-hamburger_friendship_2013}.
\revision{It is crucial for researchers to deconstruct stigmatized understandings of online friendships, which already carry the emotional burden of relying on private third spaces (\ref{subsubsec:stigma}); they do not enjoy the same privilege of public and physical availability as IRL friendships despite arguably sharing similar levels of personal depth.}



\subsubsection{Implications}
We advocate for adopting a more sensitive view of online friendship that respects their importance equally to IRL relationships. This shift would not only help alleviate stigma towards online friendships, but also support online maintenance of IRL relationships.

Currently, there is no sociotechnical infrastructure that explicitly addresses a norm difference between online and IRL friendship maintenance.
We found this manifested in practice -- users without experience on online friendship may struggle to connect when transitioning to technology platforms (\ref{subsubsec:norms_of_interaction}); this is due to the ambiguity in interaction norms on less familiar platforms.
Increasing visibility on communication platforms can help educate users about feasible norms for interaction and support more personal interactions through online media.
For example, making context cues for personal expression more visible, such as dynamic font personalization or annotating conversations with activity context cues \cite{buschek_personal_2018}, can increase visibility and awareness of a more intimate communication context~\cite{erickson2000social}.
Recent work has also shown the potential of using AI text assistants to explicitly facilitate more intimate and empathic conversations between peers \cite{chen_closer_2023,sharma_humanai_2023}.
\revision{Creating structures that facilitate better understanding and reconciliation of friendship norms across online and IRL contexts can support IRL friends transitioning to online interactions while also fostering positive recognition of online friendships. Additionally, such efforts may encourage platform strategies that lessen the need for secrecy in stigmatized online friendships.}


We also saw how the lack of groundedness in online contexts drove online friends to be creative with their appropriation of affordances (\ref{subsubsec:contextmatters}) and craft their own norms of interaction (\ref{subsubsec:norms_of_interaction}).
These could also be useful habits to cultivate for friends who are coming from more concrete norms of interaction in an IRL setting and need to maintain their relationships through technology. 
Because our findings show that “familiarity breeds contempt” \cite{norton_less_2007} in real-life relationships and leads friends to take their physical context for granted (\ref{subsubsec:norms_of_interaction}), 
designing exercises that move IRL relationships online for periods of time can be one way to promote reflection on the relationship and build a closer bond.
Encouraging activities that allow people to experience online friendships through an alternative lens can support reappraisal of stigma and even afford new types of relationship development that benefit not ‘just’ online friends but a broader scope of close relationships.

\subsection{(Strong and) Weak Ties: Supporting Friendships vs. Community Membership}
\label{subsec:strong_and_weak_ties}
Our findings add nuance to \textit{social network theory} (tie-formation) and
weak ties' relationship to strong ties.
\revision{Weak ties tend to span across more different social circles, but involve lower levels of emotional depth and closeness, while strong ties involve higher levels of closeness and trust and are likely to share the same social circles \cite{granovetter_strength_1973}.}
Despite the importance of recognizing different tie strengths \cite{gilbert_predicting_2009,vetere_mediating_2005}, much of HCI work has focused on the importance of online communities and ``the strength of weak ties'' over strong ties, which require more investment \cite{granovetter_strength_1973,zhu_impact_2014,burke_social_2011,ducheneaut_alone_2006}.
Although our work on the experiences of online friends showed that being part of a social group or community supported online friendship development, it also showed cases where the presence of weak ties in a group setting could compromise the formation of their friendships as strong ties (\ref{subsubsec:crossing_perceived_boundaries}), which are important sources of stability in people's lives (\ref{subsec:findngs_conditions}). Participant accounts showed that overwhelming group dynamics in a context with many weak ties could make the formation of strong ties counter-normative.
\revision{Specific challenges posed by large online communities to online friends included difficulty accessing personally-relevant social groups, lack of authenticity, community `drama' \cite{litt_bumpy_2014}, and 1-on-1 interactions being drowned out by community context (\ref{subsubsec:crossing_perceived_boundaries}).}
The implication that strong and weak tie formation do not always coexist in harmony suggests an air of caution when considering how well sociotechnical contexts support both types of ties.

\subsubsection{Implications}

As mentioned in the beginning of this paper, the development of high quality, strong tie friendships to maintain our well-being is a skill that our generation is currently losing.
All strong ties start out as weak; most close online friends started out meeting one another in community or group-based platform settings (e.g., interest-based forums, recreational social applications). Online friends who used individual-oriented platforms (e.g., texting apps) to maintain their relationship could only do so after establishing the friendship (see participants’ compartmentalization practices in \ref{subsubsec:crossing_perceived_boundaries}).
Hence, support for online friendship needs to start from the community-based platforms linked to its outset. 

Online friends' struggle in accessing the smaller communities within a platform where they were able to make friends (\ref{subsubsec:crossing_perceived_boundaries}) necessitates a re-examination of how we design large community platforms.
Large online social platforms (e.g., Discord servers, message boards, and VRChat) afforded the formation of smaller social groups as ‘cliques’ but did not provide much infrastructure for navigating them, which made it difficult for potential online friends to find groups where they fit in well.
One way to provide infrastructure for interpersonal relationships on community-based platforms is to integrate filtering from friend-making apps (e.g., Bumble for Friends), such as identifying personally-relevant keywords and social orientations to match online community members with socially-compatible others.
Automated matching with these characteristics for online team activities (e.g., MOBAs) can also provide opportunities to bond (\ref{subsubsec:balancing_task_personal}) with socially-compatible others if handled carefully.
Further work is needed to identify infrastructural factors relevant to online friendship development (e.g., group size, subtopics, or communication preference) and address tensions that specific interpersonal bonding could have with the platform interest.


In a similar vein, providing structured spaces in online community platforms for visible self-expression of users' individuality can also support people to find peers with a higher overlap in personal interests, which we saw as helpful to friendship development (\ref{subsubsec:balancing_task_personal}).
Additionally, creating a visible, structured space for individuality on a platform with a seemingly uniform purpose (e.g., a fandom, game, or specific activity) can not only increase the ease of finding friends but also combat strong group dynamics that might otherwise deter users from reaching out to individuals within the group (\ref{subsubsec:crossing_perceived_boundaries}).
One way to integrate such functions in an online platform may be to provide support for filtering and navigating individual users' status indicators in a centralized location. However, increasing social networking affordances may place a burden on valued anonymity, and more work is required to understand how to balance accordingly.
Appreciation of individuality can leverage existing group attributes which we found helpful for online friendship development, such as the {\small \texttt{division of labor}} (\ref{subsubsec:crossing_perceived_boundaries}) which created more opportunities for continuing interaction: increasing visibility of points of connection between users in a group can also make it easier for potential online friends to reach out to one another.

Shielding space within an online community platform for two friends to interact more privately is another way to protect friendship development originating in a highly-interactive community platform: participants mentioned how the ability to privately message one another within a community platform was drowned out if the presence of group activity was overwhelming, and online friends who were not yet close struggled to switch to a private message platform  (\ref{subsubsec:crossing_perceived_boundaries}).
Therefore, providing users with explicit options to detach their private channels from an online social platform or switch to a private platform can help developing online friends initiate the \textit{backchanneling} (\ref{subsec:internetfriends}) necessary to maintain a personal connection.

\subsection{(Time and) Space: Surpassing Media Richness With Time-Stable Interaction}
\label{subsec:time_vs_space}

Our findings identified \textit{time} as an overarching theme in the activity of online friendship.
This connects back to a diverse range of prior work on computer-mediated communication, friendship formation in MMORPGs, and {\small \texttt{activity}}.
The implications from our findings further extend upon these seminal works by exemplifying how time can be effectively considered in the design of existing and future technologies.

Section \ref{subsec:findings_modality} demonstrated how participants prioritized contextual factors (e.g., environment, current task, and time of day) instead of the sheer richness of cues (e.g., video or immersive VR) when making choices about computer-mediated communication (CMC) with their online friends.
This aligned more closely with social information processing theory (SIP) \cite{walther_social_2008} (\ref{subsec:internetfriends}) than with traditional models of media richness \cite{daft_organizational_1986}.
Online friends' co-creation of presence through their contextualized curation of technology use also connect with social presence theory \cite{biocca_toward_2003} as a fulfillment of SIP by showing how contextual considerations for the choice of CMC were made to ultimately prioritize their sense of feeling present and connected with one another.

Our findings also connect to prior work on MMORPGs which studied the development of online friendships over time. 
They aligned directly with Ramirez et al. 2018's study \cite{sukthankar_good_2018} emphasizing the importance of having different backchanneling platforms and a consistent `place to hang out' (\ref{subsec:internetfriends}). Participants reported expanding to different platforms to maintain a personal space for communicating with their friend (\ref{subsubsec:crossing_perceived_boundaries}), but also having a persistent `home' platform (\ref{subsubsec:reconcilingsynchronous}).
Participants' need to balance personal interests with social connection (\ref{subsubsec:balancing_task_personal}) also align with Utz 2000's finding on the friction between friendmaking and a meaningful activity goal \cite{utz_social_2000}.
Our work extends these findings initially observed in MMORPGs to many other contexts, such as collaborative writing, roleplaying, topic-based discussion platforms, MMOs, and other games.

Our findings on \textbf{temporal constancy} as a terminal feature of online friendship build on Nardi's \textit{dimensions of connection} for evaluating the quality of interpersonal CMC, which asserts that three relational aspects of communication (affinity, commitment, attention) form a \textit{field of connection} based on the history of communicative {\small \texttt{activity}} between two people: \textit{``The management of fields of connection requires significant interactional work to sustain communication over time.''} \cite{nardi_beyond_2005} 
Our work showcases a myriad of ways in how online friends worked to retain those \textit{dimensions of connection} over time: 
they showed commitment by crossing hurdles to personal connection (\ref{subsubsec:crossing_perceived_boundaries}), developed routines for sustained attention (\ref{subsubsec:creatingtime}), and reappropriated technologies to maximize the sense of affinity and presence felt with each other during their interactions (\ref{subsec:findings_modality}).
We not only reinforce the importance of committed contact in technology-mediated relationships, but also identify valuable examples of strategies and interactional mechanisms to support this process.

\subsubsection{Implications}
Much of the current narrative for enhancing presence in technology-mediated social interaction emphasizes spatial realism and approximating `face-to-face' contexts \cite{li_social_2021,lee_now_2011,nguyen_staying_2022,bulu_place_2012}.
But the results of our work show the stability of interaction \textit{over time} as a critical factor for cultivating strong relationships (\ref{subsubsec:creatingtime}).
\textit{Social presence theory} \cite{biocca_toward_2003} itself asserts that \textit{sense of closeness} is mediated not only by a sense of the person physically being there, but also by the level of emotional connection.
Our findings on online friends' trajectories across a variety of platforms demonstrated that dedicating time to one another over a long period was what mattered the most for keeping the friendship alive (\ref{subsec:findngs_conditions} \& \ref{subsec:findings_modality}). 
As described by our participants, the sense that their friend \textit{``had always been there''} (\ref{subsec:findngs_conditions}), \textit{``just as a person''} (\ref{subsubsec:stigma}) was not always analogous with being there physically.
Choosing less physically-immersive forms of communication and allowing room to fill in their own interpretations (\ref{subsubsec:contextmatters}) demonstrated more presence from the online friend's existence built up over time and less from direct sensory cues provided by technology.

Based on our findings, we make recommendations for designing time-stable and long-term social interactions which support friendship development. 
To enhance social presence for meaningful relationship development, future technology design should foreground the cultivation of interaction patterns that persist over time and not only immediate qualities of interaction.
As discussed in \ref{subsec:strong_and_weak_ties}, strong ties require more time commitment around which there is no shortcut.

Our findings identified several features of interaction or shared experience which were helpful for maintaining committed interaction over time, which can be combined and integrated into future systems to support friendship development.
Participating together in shared interests which involved a relatively challenging activity over time --- such as creating writing or art, moderating a server, or leveling up in a game, motivated consistent interaction by leveraging the incentive from an existing personal interest and the feeling of meaning and growth from accomplishing something difficult together (\ref{subsubsec:balancing_task_personal}). Membership in a community or group also provided more momentum for consistent dedication towards a shared activity.
To support the building of shared experiences that motivate interacting over time, future social online platforms can explicitly encourage members to connect and collaboratively act on shared interests, by not only allowing people to find each other based on interest (\ref{subsec:strong_and_weak_ties}), but also suggesting activities to pairs of potential friends and providing the connection to a space for them to do those activities together.


Additionally, making the history of \textit{meaningful} friendship activities easily visible to online friends can improve the sense of connection, presence, and motivation to sustain the friendship over time.
Participants were reassured of their friendship's existence by seeing some tangible evidence of their friendship activity, whether it was a growing village in Minecraft (\ref{subsubsec:creatingtime}), exchanged gifts (\ref{subsubsec:effortful_interaction}), or old scribbles on a virtual whiteboard (\ref{subsubsec:reconcilingsynchronous}).
Online platforms should provide more affordances for friends to more easily record, view, and interact with their activity history together, as having a tangible accumulation of friendship over time can motivate online friends to continue their connection over the virtual space. Past work has made this possible for VR experiences \cite{wang_again_2020}, but our participants reported many forms of valued activity history, ranging from collections of shared media (\ref{subsubsec:creatingtime}) to the events in different rooms of a Discord server (\ref{subsubsec:reconcilingsynchronous}).
Systems that allow retroactive asynchronous participation in usually-synchronous activities between friends, such as supporting friends to review and catch up with each other's progress in a game, can address the limitations of synchronous activities for friends with misaligned schedules.
More work is necessary to identify how to best implement the preservation and presentation of meaningful friendship activity in different platform contexts. 


\section{Limitations and Future Work}
Our study provided a rich description of online friendship development through lived experiences. But as an initial foray into understanding the online friendship process, our work still leaves many open questions to explore.
Although our participants found immense value in their friendships, more comprehensive work is needed to determine the broader impact of online friendships on topics of public concern such as loneliness and mental health.
Research moving forward should account for characteristics such as the different contexts of online and IRL friendships (\ref{subsec:contextual_imprinting}), developmental features of online friendship (e.g., turning points) (\ref{subsec:findngs_conditions}), and distinguish the online space from the friendship (\ref{subsec:strong_and_weak_ties}) --- to paint a more nuanced picture.
All of our participants had close online friendships, which may reflect survivorship bias in our findings.
Investigations that study the prevalence of online friendship across different contexts (including failed attempts), as well as its systematic relationship to online community and platform features, can provide a more complete picture of online friendship dynamics and better inform interventions to support them.

Despite our efforts to recruit online friends across a diverse range of platforms and activities (e.g., different videogames, online forums, and hobby-based platforms), most of our participants are located in the USA (72\%).
Given the culturally-contextualized nature of friendship development \cite{yum_computer-mediated_2005} and our observation of social norms' influence on online friendships, increasing representation from non-Western cultures is important for further understanding mechanisms of online friendship.

\section{Conclusion}
In this work, we performed an activity-grounded exploration to understand the distinct qualities of how online friendships develop and provide insight on how online platforms can better support this process.
Our findings reveal how online friends prioritize the use of communication technologies which sustain long-term interaction over the appeal of rich and immersive affordances.
We also extend existing theories about online socialization and technology-mediated relationship maintenance by showing how online community membership can have a complex impact on friendship development and how online friends create their own interpretations of technology use.
The current work opens a discourse for reconsideration of the value of online friendship, nurturing strong ties in online communities, and supporting time-stable interaction as a key aspect of social connection.

\bibliographystyle{ACM-Reference-Format}
\bibliography{sample-base}


\begin{thebibliography}{88}


\ifx \showCODEN    \undefined \def \showCODEN     #1{\unskip}     \fi
\ifx \showDOI      \undefined \def \showDOI       #1{#1}\fi
\ifx \showISBNx    \undefined \def \showISBNx     #1{\unskip}     \fi
\ifx \showISBNxiii \undefined \def \showISBNxiii  #1{\unskip}     \fi
\ifx \showISSN     \undefined \def \showISSN      #1{\unskip}     \fi
\ifx \showLCCN     \undefined \def \showLCCN      #1{\unskip}     \fi
\ifx \shownote     \undefined \def \shownote      #1{#1}          \fi
\ifx \showarticletitle \undefined \def \showarticletitle #1{#1}   \fi
\ifx \showURL      \undefined \def \showURL       {\relax}        \fi
\providecommand\bibfield[2]{#2}
\providecommand\bibinfo[2]{#2}
\providecommand\natexlab[1]{#1}
\providecommand\showeprint[2][]{arXiv:#2}

\bibitem[Acena and Freeman(2021)]%
        {acena_my_2021}
\bibfield{author}{\bibinfo{person}{Dane Acena} {and} \bibinfo{person}{Guo Freeman}.} \bibinfo{year}{2021}\natexlab{}.
\newblock \showarticletitle{“{In} {My} {Safe} {Space}”: {Social} {Support} for {LGBTQ} {Users} in {Social} {Virtual} {Reality}}. In \bibinfo{booktitle}{\emph{Extended {Abstracts} of the 2021 {CHI} {Conference} on {Human} {Factors} in {Computing} {Systems}}} \emph{(\bibinfo{series}{{CHI} {EA} '21})}. \bibinfo{publisher}{Association for Computing Machinery}, \bibinfo{address}{New York, NY, USA}, \bibinfo{pages}{1--6}.
\newblock
\showISBNx{978-1-4503-8095-9}
\urldef\tempurl%
\url{https://doi.org/10.1145/3411763.3451673}
\showDOI{\tempurl}


\bibitem[Altman and Taylor(1973)]%
        {altman_social_1973}
\bibfield{author}{\bibinfo{person}{Irwin Altman} {and} \bibinfo{person}{Dalmas~A. Taylor}.} \bibinfo{year}{1973}\natexlab{}.
\newblock \bibinfo{booktitle}{\emph{Social penetration: {The} development of interpersonal relationships}}.
\newblock \bibinfo{publisher}{Holt, Rinehart \& Winston}, \bibinfo{address}{Oxford, England}.
\newblock
\showISBNx{978-0-03-076635-0}
\newblock
\shownote{Pages: viii, 212}.


\bibitem[Amichai-Hamburger and Ben-Artzi(2003)]%
        {amichai-hamburger_loneliness_2003}
\bibfield{author}{\bibinfo{person}{Y Amichai-Hamburger} {and} \bibinfo{person}{E Ben-Artzi}.} \bibinfo{year}{2003}\natexlab{}.
\newblock \showarticletitle{Loneliness and {Internet} use}.
\newblock \bibinfo{journal}{\emph{Computers in Human Behavior}} \bibinfo{volume}{19}, \bibinfo{number}{1} (\bibinfo{date}{Jan.} \bibinfo{year}{2003}), \bibinfo{pages}{71--80}.
\newblock
\showISSN{0747-5632}
\urldef\tempurl%
\url{https://doi.org/10.1016/S0747-5632(02)00014-6}
\showDOI{\tempurl}


\bibitem[Amichai-Hamburger et~al\mbox{.}(2013)]%
        {amichai-hamburger_friendship_2013}
\bibfield{author}{\bibinfo{person}{Yair Amichai-Hamburger}, \bibinfo{person}{Mila Kingsbury}, {and} \bibinfo{person}{Barry~H. Schneider}.} \bibinfo{year}{2013}\natexlab{}.
\newblock \showarticletitle{Friendship: {An} old concept with a new meaning?}
\newblock \bibinfo{journal}{\emph{Computers in Human Behavior}} \bibinfo{volume}{29}, \bibinfo{number}{1} (\bibinfo{date}{Jan.} \bibinfo{year}{2013}), \bibinfo{pages}{33--39}.
\newblock
\showISSN{0747-5632}
\urldef\tempurl%
\url{https://doi.org/10.1016/j.chb.2012.05.025}
\showDOI{\tempurl}


\bibitem[Antheunis et~al\mbox{.}(2012)]%
        {antheunis_quality_2012}
\bibfield{author}{\bibinfo{person}{Marjolijn~L. Antheunis}, \bibinfo{person}{Patti~M. Valkenburg}, {and} \bibinfo{person}{Jochen Peter}.} \bibinfo{year}{2012}\natexlab{}.
\newblock \showarticletitle{The quality of online, offline, and mixed-mode friendships among users of a social networking site}.
\newblock \bibinfo{journal}{\emph{Cyberpsychology: Journal of Psychosocial Research on Cyberspace}} \bibinfo{volume}{6}, \bibinfo{number}{3} (\bibinfo{date}{Dec.} \bibinfo{year}{2012}).
\newblock
\showISSN{1802-7962}
\urldef\tempurl%
\url{https://doi.org/10.5817/CP2012-3-6}
\showDOI{\tempurl}
\newblock
\shownote{Number: 3}.


\bibitem[Arguello et~al\mbox{.}(2006)]%
        {arguello_talk_2006}
\bibfield{author}{\bibinfo{person}{Jaime Arguello}, \bibinfo{person}{Brian~S. Butler}, \bibinfo{person}{Elisabeth Joyce}, \bibinfo{person}{Robert Kraut}, \bibinfo{person}{Kimberly~S. Ling}, \bibinfo{person}{Carolyn Rosé}, {and} \bibinfo{person}{Xiaoqing Wang}.} \bibinfo{year}{2006}\natexlab{}.
\newblock \showarticletitle{Talk to me: foundations for successful individual-group interactions in online communities}. In \bibinfo{booktitle}{\emph{Proceedings of the {SIGCHI} {Conference} on {Human} {Factors} in {Computing} {Systems}}} \emph{(\bibinfo{series}{{CHI} '06})}. \bibinfo{publisher}{Association for Computing Machinery}, \bibinfo{address}{New York, NY, USA}, \bibinfo{pages}{959--968}.
\newblock
\showISBNx{978-1-59593-372-0}
\urldef\tempurl%
\url{https://doi.org/10.1145/1124772.1124916}
\showDOI{\tempurl}


\bibitem[Armstrong(2022)]%
        {armstrong_friendships_2022}
\bibfield{author}{\bibinfo{person}{Martin Armstrong}.} \bibinfo{year}{2022}\natexlab{}.
\newblock \bibinfo{title}{Friendships: {Less} is now more}.
\newblock
\newblock
\urldef\tempurl%
\url{https://www.weforum.org/agenda/2022/11/friendships-less-is-now-more/}
\showURL{%
\tempurl}


\bibitem[Austin et~al\mbox{.}(2020)]%
        {austin_its_2020}
\bibfield{author}{\bibinfo{person}{Ashley Austin}, \bibinfo{person}{Shelley~L. Craig}, \bibinfo{person}{Nicole Navega}, {and} \bibinfo{person}{Lauren~B. McInroy}.} \bibinfo{year}{2020}\natexlab{}.
\newblock \showarticletitle{It’s my safe space: {The} life-saving role of the internet in the lives of transgender and gender diverse youth}.
\newblock \bibinfo{journal}{\emph{International Journal of Transgender Health}} \bibinfo{volume}{21}, \bibinfo{number}{1} (\bibinfo{date}{Jan.} \bibinfo{year}{2020}), \bibinfo{pages}{33--44}.
\newblock
\showISSN{2689-5269}
\urldef\tempurl%
\url{https://doi.org/10.1080/15532739.2019.1700202}
\showDOI{\tempurl}
\newblock
\shownote{Publisher: Taylor \& Francis \_eprint: https://doi.org/10.1080/15532739.2019.1700202}.


\bibitem[Baym(2010)]%
        {baym_personal_2010}
\bibfield{author}{\bibinfo{person}{Nancy~K. Baym}.} \bibinfo{year}{2010}\natexlab{}.
\newblock \bibinfo{booktitle}{\emph{Personal connections in the digital age}}.
\newblock \bibinfo{publisher}{Polity}, \bibinfo{address}{Cambridge, UK}.
\newblock
\showISBNx{978-0-7456-4331-1}
\urldef\tempurl%
\url{https://shibboleth2sp.gar.semcs.net/Shibboleth.sso/Login?entityID=https://passport01.leeds.ac.uk/idp/shibboleth&target=https://shibboleth2sp.gar.semcs.net/shib?dest=http%253A%252F%252Fwww.vlebooks.com%252FSHIBBOLETH%253Fdest%253Dhttp%25253A%25252F%25252Fwww.vlebooks.com%25252Fvleweb%25252Fproduct%25252Fopenreader%25253Fid%25253DLeedsUni%252526isbn%25253D9780745656199}
\showURL{%
\tempurl}
\newblock
\shownote{OCLC: 499095289}.


\bibitem[Biocca et~al\mbox{.}(2003)]%
        {biocca_toward_2003}
\bibfield{author}{\bibinfo{person}{Frank Biocca}, \bibinfo{person}{Chad Harms}, {and} \bibinfo{person}{Judee~K. Burgoon}.} \bibinfo{year}{2003}\natexlab{}.
\newblock \showarticletitle{Toward a {More} {Robust} {Theory} and {Measure} of {Social} {Presence}: {Review} and {Suggested} {Criteria}}.
\newblock \bibinfo{journal}{\emph{Presence: Teleoperators and Virtual Environments}} \bibinfo{volume}{12}, \bibinfo{number}{5} (\bibinfo{date}{Oct.} \bibinfo{year}{2003}), \bibinfo{pages}{456--480}.
\newblock
\urldef\tempurl%
\url{https://doi.org/10.1162/105474603322761270}
\showDOI{\tempurl}


\bibitem[Briggle(2008)]%
        {briggle_real_2008}
\bibfield{author}{\bibinfo{person}{Adam Briggle}.} \bibinfo{year}{2008}\natexlab{}.
\newblock \showarticletitle{Real friends: how the {Internet} can foster friendship}.
\newblock \bibinfo{journal}{\emph{Ethics and Information Technology}} \bibinfo{volume}{10}, \bibinfo{number}{1} (\bibinfo{date}{March} \bibinfo{year}{2008}), \bibinfo{pages}{71--79}.
\newblock
\showISSN{1572-8439}
\urldef\tempurl%
\url{https://doi.org/10.1007/s10676-008-9160-z}
\showDOI{\tempurl}


\bibitem[Bulu(2012)]%
        {bulu_place_2012}
\bibfield{author}{\bibinfo{person}{Saniye~Tugba Bulu}.} \bibinfo{year}{2012}\natexlab{}.
\newblock \showarticletitle{Place presence, social presence, co-presence, and satisfaction in virtual worlds}.
\newblock \bibinfo{journal}{\emph{Computers \& Education}} \bibinfo{volume}{58}, \bibinfo{number}{1} (\bibinfo{date}{Jan.} \bibinfo{year}{2012}), \bibinfo{pages}{154--161}.
\newblock
\showISSN{0360-1315}
\urldef\tempurl%
\url{https://doi.org/10.1016/j.compedu.2011.08.024}
\showDOI{\tempurl}


\bibitem[Burke et~al\mbox{.}(2011)]%
        {burke_social_2011}
\bibfield{author}{\bibinfo{person}{Moira Burke}, \bibinfo{person}{Robert Kraut}, {and} \bibinfo{person}{Cameron Marlow}.} \bibinfo{year}{2011}\natexlab{}.
\newblock \showarticletitle{Social capital on facebook: differentiating uses and users}. In \bibinfo{booktitle}{\emph{Proceedings of the {SIGCHI} {Conference} on {Human} {Factors} in {Computing} {Systems}}} \emph{(\bibinfo{series}{{CHI} '11})}. \bibinfo{publisher}{Association for Computing Machinery}, \bibinfo{address}{New York, NY, USA}, \bibinfo{pages}{571--580}.
\newblock
\showISBNx{978-1-4503-0228-9}
\urldef\tempurl%
\url{https://doi.org/10.1145/1978942.1979023}
\showDOI{\tempurl}


\bibitem[Buschek et~al\mbox{.}(2018)]%
        {buschek_personal_2018}
\bibfield{author}{\bibinfo{person}{Daniel Buschek}, \bibinfo{person}{Mariam Hassib}, {and} \bibinfo{person}{Florian Alt}.} \bibinfo{year}{2018}\natexlab{}.
\newblock \showarticletitle{Personal {Mobile} {Messaging} in {Context}: {Chat} {Augmentations} for {Expressiveness} and {Awareness}}.
\newblock \bibinfo{journal}{\emph{ACM Trans. Comput.-Hum. Interact.}} \bibinfo{volume}{25}, \bibinfo{number}{4} (\bibinfo{date}{Aug.} \bibinfo{year}{2018}), \bibinfo{pages}{23:1--23:33}.
\newblock
\showISSN{1073-0516}
\urldef\tempurl%
\url{https://doi.org/10.1145/3201404}
\showDOI{\tempurl}


\bibitem[Cairns et~al\mbox{.}(2020)]%
        {cairns_covid-19_2020}
\bibfield{author}{\bibinfo{person}{Maryann~R. Cairns}, \bibinfo{person}{Margaret Ebinger}, \bibinfo{person}{Chanel Stinson}, {and} \bibinfo{person}{Jason Jordan}.} \bibinfo{year}{2020}\natexlab{}.
\newblock \showarticletitle{{COVID}-19 and {Human} {Connection}: {Collaborative} {Research} on {Loneliness} and {Online} {Worlds} from a {Socially}-{Distanced} {Academy}}.
\newblock \bibinfo{journal}{\emph{Human Organization}} \bibinfo{volume}{79}, \bibinfo{number}{4} (\bibinfo{date}{Dec.} \bibinfo{year}{2020}), \bibinfo{pages}{281--291}.
\newblock
\showISSN{0018-7259}
\urldef\tempurl%
\url{https://doi.org/10.17730/1938-3525-79.4.281}
\showDOI{\tempurl}


\bibitem[Chan and Cheng(2004)]%
        {chan_comparison_2004}
\bibfield{author}{\bibinfo{person}{Darius K.-S. Chan} {and} \bibinfo{person}{Grand H.-L. Cheng}.} \bibinfo{year}{2004}\natexlab{}.
\newblock \showarticletitle{A {Comparison} of {Offline} and {Online} {Friendship} {Qualities} at {Different} {Stages} of {Relationship} {Development}}.
\newblock \bibinfo{journal}{\emph{Journal of Social and Personal Relationships}} \bibinfo{volume}{21}, \bibinfo{number}{3} (\bibinfo{date}{June} \bibinfo{year}{2004}), \bibinfo{pages}{305--320}.
\newblock
\showISSN{0265-4075}
\urldef\tempurl%
\url{https://doi.org/10.1177/0265407504042834}
\showDOI{\tempurl}
\newblock
\shownote{Publisher: SAGE Publications Ltd}.


\bibitem[Charmaz(2006)]%
        {charmaz_constructing_2006}
\bibfield{author}{\bibinfo{person}{Kathy Charmaz}.} \bibinfo{year}{2006}\natexlab{}.
\newblock \bibinfo{booktitle}{\emph{Constructing grounded theory: {A} practical guide through qualitative analysis}}.
\newblock \bibinfo{publisher}{SAGE Publications}, \bibinfo{address}{London}.
\newblock
\urldef\tempurl%
\url{https://books.google.com/books?hl=en&lr=&id=2ThdBAAAQBAJ&oi=fnd&pg=PP1&dq=Charmaz,+K.+(2006).+Constructing+grounded+theory:+A+practical+guide+through+qualitative+analysis.+London:+Sage.&ots=f-jTbOrDyX&sig=JWXxnn9R0hEsOrsMxVjkWnPNNIE}
\showURL{%
\tempurl}


\bibitem[Chayka(2015)]%
        {chayka_lets_2015}
\bibfield{author}{\bibinfo{person}{Kyle Chayka}.} \bibinfo{year}{2015}\natexlab{}.
\newblock \showarticletitle{Let’s {Really} {Be} {Friends}: {A} defense of online intimacy}.
\newblock \bibinfo{journal}{\emph{The New Republic}} (\bibinfo{date}{March} \bibinfo{year}{2015}).
\newblock
\showISSN{0028-6583}
\urldef\tempurl%
\url{https://newrepublic.com/article/121183/your-internet-friends-are-real-defense-online-intimacy}
\showURL{%
\tempurl}


\bibitem[Chayko(2020)]%
        {chayko_superconnected_2020}
\bibfield{author}{\bibinfo{person}{Mary Chayko}.} \bibinfo{year}{2020}\natexlab{}.
\newblock \bibinfo{booktitle}{\emph{Superconnected: {The} {Internet}, {Digital} {Media}, and {Techno}-{Social} {Life}}}.
\newblock \bibinfo{publisher}{SAGE Publications}.
\newblock
\showISBNx{978-1-07-180528-2}
\newblock
\shownote{Google-Books-ID: JJYFEAAAQBAJ}.


\bibitem[Chen et~al\mbox{.}(2023)]%
        {chen_closer_2023}
\bibfield{author}{\bibinfo{person}{Tiffany Chen}, \bibinfo{person}{Cassandra Lee}, \bibinfo{person}{Jessica~R Mindel}, \bibinfo{person}{Neska Elhaouij}, {and} \bibinfo{person}{Rosalind Picard}.} \bibinfo{year}{2023}\natexlab{}.
\newblock \showarticletitle{Closer {Worlds}: {Using} {Generative} {AI} to {Facilitate} {Intimate} {Conversations}}. In \bibinfo{booktitle}{\emph{Extended {Abstracts} of the 2023 {CHI} {Conference} on {Human} {Factors} in {Computing} {Systems}}} \emph{(\bibinfo{series}{{CHI} {EA} '23})}. \bibinfo{publisher}{Association for Computing Machinery}, \bibinfo{address}{New York, NY, USA}, \bibinfo{pages}{1--15}.
\newblock
\showISBNx{978-1-4503-9422-2}
\urldef\tempurl%
\url{https://doi.org/10.1145/3544549.3585651}
\showDOI{\tempurl}


\bibitem[Cox(2021)]%
        {cox_state_2021}
\bibfield{author}{\bibinfo{person}{Daniel~A. Cox}.} \bibinfo{year}{2021}\natexlab{}.
\newblock \bibinfo{booktitle}{\emph{The {State} of {American} {Friendship}: {Change}, {Challenges}, and {Loss}}}.
\newblock \bibinfo{type}{{T}echnical {R}eport}. \bibinfo{institution}{Survey Center on American Life}.
\newblock
\urldef\tempurl%
\url{https://www.americansurveycenter.org/research/the-state-of-american-friendship-change-challenges-and-loss/}
\showURL{%
\tempurl}


\bibitem[Daft and Lengel(1986)]%
        {daft_organizational_1986}
\bibfield{author}{\bibinfo{person}{Richard~L. Daft} {and} \bibinfo{person}{Robert~H. Lengel}.} \bibinfo{year}{1986}\natexlab{}.
\newblock \showarticletitle{Organizational {Information} {Requirements}, {Media} {Richness} and {Structural} {Design}}.
\newblock \bibinfo{journal}{\emph{Management Science}} \bibinfo{volume}{32}, \bibinfo{number}{5} (\bibinfo{date}{May} \bibinfo{year}{1986}), \bibinfo{pages}{554--571}.
\newblock
\showISSN{0025-1909}
\urldef\tempurl%
\url{https://doi.org/10.1287/mnsc.32.5.554}
\showDOI{\tempurl}
\newblock
\shownote{Publisher: INFORMS}.


\bibitem[Davis and Todd(1982)]%
        {davis_friendship_1982}
\bibfield{author}{\bibinfo{person}{Keith~E. Davis} {and} \bibinfo{person}{Michael~J. Todd}.} \bibinfo{year}{1982}\natexlab{}.
\newblock \showarticletitle{Friendship and love relationships}.
\newblock \bibinfo{journal}{\emph{Advances in Descriptive Psychology}}  \bibinfo{volume}{2} (\bibinfo{year}{1982}), \bibinfo{pages}{79--122}.
\newblock
\showISSN{0276-9913}
\newblock
\shownote{Place: US Publisher: JAI Press, Inc.}.


\bibitem[Deighan et~al\mbox{.}(2023)]%
        {deighan_social_2023}
\bibfield{author}{\bibinfo{person}{Mairi~Therese Deighan}, \bibinfo{person}{Amid Ayobi}, {and} \bibinfo{person}{Aisling~Ann O'Kane}.} \bibinfo{year}{2023}\natexlab{}.
\newblock \showarticletitle{Social {Virtual} {Reality} as a {Mental} {Health} {Tool}: {How} {People} {Use} {VRChat} to {Support} {Social} {Connectedness} and {Wellbeing}}. In \bibinfo{booktitle}{\emph{Proceedings of the 2023 {CHI} {Conference} on {Human} {Factors} in {Computing} {Systems}}} \emph{(\bibinfo{series}{{CHI} '23})}. \bibinfo{publisher}{Association for Computing Machinery}, \bibinfo{address}{New York, NY, USA}, \bibinfo{pages}{1--13}.
\newblock
\showISBNx{978-1-4503-9421-5}
\urldef\tempurl%
\url{https://doi.org/10.1145/3544548.3581103}
\showDOI{\tempurl}


\bibitem[Dinic(2021)]%
        {dinic_yougov_2021}
\bibfield{author}{\bibinfo{person}{Milan Dinic}.} \bibinfo{year}{2021}\natexlab{}.
\newblock \bibinfo{booktitle}{\emph{The {YouGov} {Friendship} {Study} {\textbar} {YouGov}}}.
\newblock \bibinfo{type}{{T}echnical {R}eport}. \bibinfo{institution}{YouGov}.
\newblock
\urldef\tempurl%
\url{https://yougov.co.uk/society/articles/38491-yougov-friendship-study}
\showURL{%
\tempurl}


\bibitem[Ducheneaut et~al\mbox{.}(2006)]%
        {ducheneaut_alone_2006}
\bibfield{author}{\bibinfo{person}{Nicolas Ducheneaut}, \bibinfo{person}{Nicholas Yee}, \bibinfo{person}{Eric Nickell}, {and} \bibinfo{person}{Robert~J. Moore}.} \bibinfo{year}{2006}\natexlab{}.
\newblock \showarticletitle{"{Alone} together?": exploring the social dynamics of massively multiplayer online games}. In \bibinfo{booktitle}{\emph{Proceedings of the {SIGCHI} {Conference} on {Human} {Factors} in {Computing} {Systems}}} \emph{(\bibinfo{series}{{CHI} '06})}. \bibinfo{publisher}{Association for Computing Machinery}, \bibinfo{address}{New York, NY, USA}, \bibinfo{pages}{407--416}.
\newblock
\showISBNx{978-1-59593-372-0}
\urldef\tempurl%
\url{https://doi.org/10.1145/1124772.1124834}
\showDOI{\tempurl}


\bibitem[Engeström(2014)]%
        {engestrom_learning_2014}
\bibfield{author}{\bibinfo{person}{Yrjö Engeström}.} \bibinfo{year}{2014}\natexlab{}.
\newblock \bibinfo{booktitle}{\emph{Learning by {Expanding}: {An} {Activity}-{Theoretical} {Approach} to {Developmental} {Research}} (\bibinfo{edition}{2} ed.)}.
\newblock \bibinfo{publisher}{Cambridge University Press}, \bibinfo{address}{Cambridge}.
\newblock
\showISBNx{978-1-107-07442-2}
\urldef\tempurl%
\url{https://doi.org/10.1017/CBO9781139814744}
\showDOI{\tempurl}


\bibitem[Erickson and Kellogg(2000)]%
        {erickson2000social}
\bibfield{author}{\bibinfo{person}{Thomas Erickson} {and} \bibinfo{person}{Wendy~A Kellogg}.} \bibinfo{year}{2000}\natexlab{}.
\newblock \showarticletitle{Social translucence: an approach to designing systems that support social processes}.
\newblock \bibinfo{journal}{\emph{ACM transactions on computer-human interaction (TOCHI)}} \bibinfo{volume}{7}, \bibinfo{number}{1} (\bibinfo{year}{2000}), \bibinfo{pages}{59--83}.
\newblock


\bibitem[Fehr(1996)]%
        {fehr_friendship_1996}
\bibfield{author}{\bibinfo{person}{Beverley Fehr}.} \bibinfo{year}{1996}\natexlab{}.
\newblock \bibinfo{booktitle}{\emph{Friendship processes}}.
\newblock \bibinfo{publisher}{Sage Publications}, \bibinfo{address}{Thousand Oaks, California}.
\newblock
\showISBNx{978-0-8039-4560-9}
\urldef\tempurl%
\url{http://catdir.loc.gov/catdir/enhancements/fy0655/95032479-t.html}
\showURL{%
\tempurl}
\newblock
\shownote{OCLC: 32924872}.


\bibitem[Freeman et~al\mbox{.}(2016)]%
        {freeman_revisiting_2016}
\bibfield{author}{\bibinfo{person}{Guo Freeman}, \bibinfo{person}{Jeffrey Bardzell}, {and} \bibinfo{person}{Shaowen Bardzell}.} \bibinfo{year}{2016}\natexlab{}.
\newblock \showarticletitle{Revisiting {Computer}-{Mediated} {Intimacy}: {In}-{Game} {Marriage} and {Dyadic} {Gameplay} in {Audition}}. In \bibinfo{booktitle}{\emph{Proceedings of the 2016 {CHI} {Conference} on {Human} {Factors} in {Computing} {Systems}}} \emph{(\bibinfo{series}{{CHI} '16})}. \bibinfo{publisher}{Association for Computing Machinery}, \bibinfo{address}{New York, NY, USA}, \bibinfo{pages}{4325--4336}.
\newblock
\showISBNx{978-1-4503-3362-7}
\urldef\tempurl%
\url{https://doi.org/10.1145/2858036.2858484}
\showDOI{\tempurl}


\bibitem[Fuhrman et~al\mbox{.}(2009)]%
        {fuhrman_behavior_2009}
\bibfield{author}{\bibinfo{person}{Robert~W. Fuhrman}, \bibinfo{person}{Dorothy Flannagan}, {and} \bibinfo{person}{Mike Matamoros}.} \bibinfo{year}{2009}\natexlab{}.
\newblock \showarticletitle{Behavior expectations in cross-sex friendships, same-sex friendships, and romantic relationships}.
\newblock \bibinfo{journal}{\emph{Personal Relationships}} \bibinfo{volume}{16}, \bibinfo{number}{4} (\bibinfo{year}{2009}), \bibinfo{pages}{575--596}.
\newblock
\showISSN{1475-6811}
\urldef\tempurl%
\url{https://doi.org/10.1111/j.1475-6811.2009.01240.x}
\showDOI{\tempurl}
\newblock
\shownote{\_eprint: https://onlinelibrary.wiley.com/doi/pdf/10.1111/j.1475-6811.2009.01240.x}.


\bibitem[Gatos et~al\mbox{.}(2021)]%
        {gatos_how_2021}
\bibfield{author}{\bibinfo{person}{Doğa Gatos}, \bibinfo{person}{Aslı Günay}, \bibinfo{person}{Güncel Kırlangıç}, \bibinfo{person}{Kemal Kuscu}, {and} \bibinfo{person}{Asim~Evren Yantac}.} \bibinfo{year}{2021}\natexlab{}.
\newblock \showarticletitle{How {HCI} {Bridges} {Health} and {Design} in {Online} {Health} {Communities}: {A} {Systematic} {Review}}. In \bibinfo{booktitle}{\emph{Proceedings of the 2021 {ACM} {Designing} {Interactive} {Systems} {Conference}}} \emph{(\bibinfo{series}{{DIS} '21})}. \bibinfo{publisher}{Association for Computing Machinery}, \bibinfo{address}{New York, NY, USA}, \bibinfo{pages}{970--983}.
\newblock
\showISBNx{978-1-4503-8476-6}
\urldef\tempurl%
\url{https://doi.org/10.1145/3461778.3462100}
\showDOI{\tempurl}


\bibitem[Ghosh et~al\mbox{.}(2023)]%
        {ghosh_i_2023}
\bibfield{author}{\bibinfo{person}{Sourojit Ghosh}, \bibinfo{person}{Niamh Froelich}, {and} \bibinfo{person}{Cecilia Aragon}.} \bibinfo{year}{2023}\natexlab{}.
\newblock \showarticletitle{“{I} {Love} {You}, {My} {Dear} {Friend}”: {Analyzing} the {Role} of {Emotions} in the {Building} of {Friendships} in {Online} {Fanfiction} {Communities}}. In \bibinfo{booktitle}{\emph{Social {Computing} and {Social} {Media}}}, \bibfield{editor}{\bibinfo{person}{Adela Coman} {and} \bibinfo{person}{Simona Vasilache}} (Eds.). \bibinfo{publisher}{Springer Nature Switzerland}, \bibinfo{address}{Cham}, \bibinfo{pages}{466--485}.
\newblock
\showISBNx{978-3-031-35927-9}
\urldef\tempurl%
\url{https://doi.org/10.1007/978-3-031-35927-9_32}
\showDOI{\tempurl}


\bibitem[Gilbert and Karahalios(2009)]%
        {gilbert_predicting_2009}
\bibfield{author}{\bibinfo{person}{Eric Gilbert} {and} \bibinfo{person}{Karrie Karahalios}.} \bibinfo{year}{2009}\natexlab{}.
\newblock \showarticletitle{Predicting tie strength with social media}. In \bibinfo{booktitle}{\emph{Proceedings of the {SIGCHI} {Conference} on {Human} {Factors} in {Computing} {Systems}}} \emph{(\bibinfo{series}{{CHI} '09})}. \bibinfo{publisher}{Association for Computing Machinery}, \bibinfo{address}{New York, NY, USA}, \bibinfo{pages}{211--220}.
\newblock
\showISBNx{978-1-60558-246-7}
\urldef\tempurl%
\url{https://doi.org/10.1145/1518701.1518736}
\showDOI{\tempurl}


\bibitem[Granovetter(1973)]%
        {granovetter_strength_1973}
\bibfield{author}{\bibinfo{person}{Mark~S. Granovetter}.} \bibinfo{year}{1973}\natexlab{}.
\newblock \showarticletitle{The {Strength} of {Weak} {Ties}}.
\newblock \bibinfo{journal}{\emph{Amer. J. Sociology}} \bibinfo{volume}{78}, \bibinfo{number}{6} (\bibinfo{date}{May} \bibinfo{year}{1973}), \bibinfo{pages}{1360--1380}.
\newblock
\showISSN{0002-9602}
\urldef\tempurl%
\url{https://doi.org/10.1086/225469}
\showDOI{\tempurl}
\newblock
\shownote{Publisher: The University of Chicago Press}.


\bibitem[Han et~al\mbox{.}(2019)]%
        {han_what_2019}
\bibfield{author}{\bibinfo{person}{Xi Han}, \bibinfo{person}{Wenting Han}, \bibinfo{person}{Jiabin Qu}, \bibinfo{person}{Bei Li}, {and} \bibinfo{person}{Qinghua Zhu}.} \bibinfo{year}{2019}\natexlab{}.
\newblock \showarticletitle{What happens online stays online? —— {Social} media dependency, online support behavior and offline effects for {LGBT}}.
\newblock \bibinfo{journal}{\emph{Computers in Human Behavior}}  \bibinfo{volume}{93} (\bibinfo{date}{April} \bibinfo{year}{2019}), \bibinfo{pages}{91--98}.
\newblock
\showISSN{0747-5632}
\urldef\tempurl%
\url{https://doi.org/10.1016/j.chb.2018.12.011}
\showDOI{\tempurl}


\bibitem[Hays(1984)]%
        {hays_development_1984}
\bibfield{author}{\bibinfo{person}{Robert~B. Hays}.} \bibinfo{year}{1984}\natexlab{}.
\newblock \showarticletitle{The {Development} and {Maintenance} of {Friendship}}.
\newblock \bibinfo{journal}{\emph{Journal of Social and Personal Relationships}} \bibinfo{volume}{1}, \bibinfo{number}{1} (\bibinfo{date}{March} \bibinfo{year}{1984}), \bibinfo{pages}{75--98}.
\newblock
\showISSN{0265-4075}
\urldef\tempurl%
\url{https://doi.org/10.1177/0265407584011005}
\showDOI{\tempurl}
\newblock
\shownote{Publisher: SAGE Publications Ltd}.


\bibitem[Hillier et~al\mbox{.}(2012)]%
        {hillier_internet_2012}
\bibfield{author}{\bibinfo{person}{Lynne Hillier}, \bibinfo{person}{Kimberly~J. Mitchell}, {and} \bibinfo{person}{Michele~L. Ybarra}.} \bibinfo{year}{2012}\natexlab{}.
\newblock \showarticletitle{The {Internet} {As} a {Safety} {Net}: {Findings} {From} a {Series} of {Online} {Focus} {Groups} {With} {LGB} and {Non}-{LGB} {Young} {People} in the {United} {States}}.
\newblock \bibinfo{journal}{\emph{Journal of LGBT Youth}} \bibinfo{volume}{9}, \bibinfo{number}{3} (\bibinfo{date}{July} \bibinfo{year}{2012}), \bibinfo{pages}{225--246}.
\newblock
\showISSN{1936-1653}
\urldef\tempurl%
\url{https://doi.org/10.1080/19361653.2012.684642}
\showDOI{\tempurl}
\newblock
\shownote{Publisher: Routledge \_eprint: https://doi.org/10.1080/19361653.2012.684642}.


\bibitem[Johnson and Weller(2002)]%
        {johnson2002elicitation}
\bibfield{author}{\bibinfo{person}{Jeffrey~C Johnson} {and} \bibinfo{person}{Susan~C Weller}.} \bibinfo{year}{2002}\natexlab{}.
\newblock \showarticletitle{Elicitation techniques for interviewing}.
\newblock \bibinfo{journal}{\emph{Handbook of interview research: Context and method}} (\bibinfo{year}{2002}), \bibinfo{pages}{491--514}.
\newblock


\bibitem[Kaptelinin(1995)]%
        {kaptelinin_computer-mediated_1995}
\bibfield{author}{\bibinfo{person}{Victor Kaptelinin}.} \bibinfo{year}{1995}\natexlab{}.
\newblock \showarticletitle{Computer-mediated activity: functional organs in social and developmental contexts}.
\newblock In \bibinfo{booktitle}{\emph{Context and consciousness: activity theory and human-computer interaction}}. \bibinfo{publisher}{Massachusetts Institute of Technology}, \bibinfo{address}{USA}, \bibinfo{pages}{45--68}.
\newblock
\showISBNx{978-0-262-14058-4}


\bibitem[Kozulin(1990)]%
        {kozulin_vygotskys_1990}
\bibfield{author}{\bibinfo{person}{Alex Kozulin}.} \bibinfo{year}{1990}\natexlab{}.
\newblock \bibinfo{booktitle}{\emph{Vygotsky's psychology: {A} biography of ideas}}.
\newblock \bibinfo{publisher}{Harvard University Press}, \bibinfo{address}{Cambridge, MA, US}.
\newblock
\showISBNx{978-0-674-94365-0}
\newblock
\shownote{Pages: 286}.


\bibitem[Kuutti(1995)]%
        {kuutti_activity_1995}
\bibfield{author}{\bibinfo{person}{Kari Kuutti}.} \bibinfo{year}{1995}\natexlab{}.
\newblock \showarticletitle{Activity theory as a potential framework for human-computer interaction research}.
\newblock In \bibinfo{booktitle}{\emph{Context and consciousness: activity theory and human-computer interaction}}. \bibinfo{publisher}{Massachusetts Institute of Technology}, \bibinfo{address}{USA}, \bibinfo{pages}{17--44}.
\newblock
\showISBNx{978-0-262-14058-4}


\bibitem[Lai and Fung(2020)]%
        {lai_online_2020}
\bibfield{author}{\bibinfo{person}{Gina Lai} {and} \bibinfo{person}{Ka~Yi Fung}.} \bibinfo{year}{2020}\natexlab{}.
\newblock \showarticletitle{From online strangers to offline friends: a qualitative study of video game players in {Hong} {Kong}}.
\newblock \bibinfo{journal}{\emph{Media, Culture \& Society}} \bibinfo{volume}{42}, \bibinfo{number}{4} (\bibinfo{date}{May} \bibinfo{year}{2020}), \bibinfo{pages}{483--501}.
\newblock
\showISSN{0163-4437}
\urldef\tempurl%
\url{https://doi.org/10.1177/0163443719853505}
\showDOI{\tempurl}
\newblock
\shownote{Publisher: SAGE Publications Ltd}.


\bibitem[Lampe et~al\mbox{.}(2010)]%
        {lampe_motivations_2010}
\bibfield{author}{\bibinfo{person}{Cliff Lampe}, \bibinfo{person}{Rick Wash}, \bibinfo{person}{Alcides Velasquez}, {and} \bibinfo{person}{Elif Ozkaya}.} \bibinfo{year}{2010}\natexlab{}.
\newblock \showarticletitle{Motivations to participate in online communities}. In \bibinfo{booktitle}{\emph{Proceedings of the {SIGCHI} {Conference} on {Human} {Factors} in {Computing} {Systems}}} \emph{(\bibinfo{series}{{CHI} '10})}. \bibinfo{publisher}{Association for Computing Machinery}, \bibinfo{address}{New York, NY, USA}, \bibinfo{pages}{1927--1936}.
\newblock
\showISBNx{978-1-60558-929-9}
\urldef\tempurl%
\url{https://doi.org/10.1145/1753326.1753616}
\showDOI{\tempurl}


\bibitem[Lampe et~al\mbox{.}(2007)]%
        {lampe_familiar_2007}
\bibfield{author}{\bibinfo{person}{Cliff~A.C. Lampe}, \bibinfo{person}{Nicole Ellison}, {and} \bibinfo{person}{Charles Steinfield}.} \bibinfo{year}{2007}\natexlab{}.
\newblock \showarticletitle{A familiar face(book): profile elements as signals in an online social network}. In \bibinfo{booktitle}{\emph{Proceedings of the {SIGCHI} {Conference} on {Human} {Factors} in {Computing} {Systems}}} \emph{(\bibinfo{series}{{CHI} '07})}. \bibinfo{publisher}{Association for Computing Machinery}, \bibinfo{address}{New York, NY, USA}, \bibinfo{pages}{435--444}.
\newblock
\showISBNx{978-1-59593-593-9}
\urldef\tempurl%
\url{https://doi.org/10.1145/1240624.1240695}
\showDOI{\tempurl}


\bibitem[Larivière-Bastien et~al\mbox{.}(2022)]%
        {lariviere-bastien_childrens_2022}
\bibfield{author}{\bibinfo{person}{Danaë Larivière-Bastien}, \bibinfo{person}{Olivier Aubuchon}, \bibinfo{person}{Aurélie Blondin}, \bibinfo{person}{Dominique Dupont}, \bibinfo{person}{Jamie Libenstein}, \bibinfo{person}{Florence Séguin}, \bibinfo{person}{Alexandra Tremblay}, \bibinfo{person}{Hamza Zarglayoun}, \bibinfo{person}{Catherine~M. Herba}, {and} \bibinfo{person}{Miriam~H. Beauchamp}.} \bibinfo{year}{2022}\natexlab{}.
\newblock \showarticletitle{Children's perspectives on friendships and socialization during the {COVID}-19 pandemic: {A} qualitative approach}.
\newblock \bibinfo{journal}{\emph{Child: Care, Health and Development}} \bibinfo{volume}{48}, \bibinfo{number}{6} (\bibinfo{year}{2022}), \bibinfo{pages}{1017--1030}.
\newblock
\showISSN{1365-2214}
\urldef\tempurl%
\url{https://doi.org/10.1111/cch.12998}
\showDOI{\tempurl}
\newblock
\shownote{\_eprint: https://onlinelibrary.wiley.com/doi/pdf/10.1111/cch.12998}.


\bibitem[Lee and Takayama(2011)]%
        {lee_now_2011}
\bibfield{author}{\bibinfo{person}{Min~Kyung Lee} {and} \bibinfo{person}{Leila Takayama}.} \bibinfo{year}{2011}\natexlab{}.
\newblock \showarticletitle{"{Now}, i have a body": uses and social norms for mobile remote presence in the workplace}. In \bibinfo{booktitle}{\emph{Proceedings of the {SIGCHI} {Conference} on {Human} {Factors} in {Computing} {Systems}}} \emph{(\bibinfo{series}{{CHI} '11})}. \bibinfo{publisher}{Association for Computing Machinery}, \bibinfo{address}{New York, NY, USA}, \bibinfo{pages}{33--42}.
\newblock
\showISBNx{978-1-4503-0228-9}
\urldef\tempurl%
\url{https://doi.org/10.1145/1978942.1978950}
\showDOI{\tempurl}


\bibitem[Lee and Toyama(2024)]%
        {lee_friendship_2024}
\bibfield{author}{\bibinfo{person}{Soyoung Lee} {and} \bibinfo{person}{Kentaro Toyama}.} \bibinfo{year}{2024}\natexlab{}.
\newblock \showarticletitle{Friendship {Formation} in an {Enforced} {Online} {Regime}: {Findings} from a {U}.{S}. {University} {Under} {COVID}}.
\newblock \bibinfo{journal}{\emph{Proc. ACM Hum.-Comput. Interact.}} \bibinfo{volume}{8}, \bibinfo{number}{CSCW1} (\bibinfo{date}{April} \bibinfo{year}{2024}), \bibinfo{pages}{168:1--168:30}.
\newblock
\urldef\tempurl%
\url{https://doi.org/10.1145/3641007}
\showDOI{\tempurl}


\bibitem[Leont'ev(1978)]%
        {leontev_activity_1978}
\bibfield{author}{\bibinfo{person}{A.~N. (Aleksei~Nikolaevich) Leont'ev}.} \bibinfo{year}{1978}\natexlab{}.
\newblock \bibinfo{booktitle}{\emph{Activity, consciousness, and personality}}.
\newblock \bibinfo{publisher}{Prentice-Hall}, \bibinfo{address}{Englewood Cliffs, N.J.}
\newblock
\urldef\tempurl%
\url{https://search.library.wisc.edu/catalog/999498791102121}
\showURL{%
\tempurl}


\bibitem[Li et~al\mbox{.}(2021)]%
        {li_social_2021}
\bibfield{author}{\bibinfo{person}{Jie Li}, \bibinfo{person}{Vinoba Vinayagamoorthy}, \bibinfo{person}{Julie Williamson}, \bibinfo{person}{David~A. Shamma}, {and} \bibinfo{person}{Pablo Cesar}.} \bibinfo{year}{2021}\natexlab{}.
\newblock \showarticletitle{Social {VR}: {A} {New} {Medium} for {Remote} {Communication} and {Collaboration}}. In \bibinfo{booktitle}{\emph{Extended {Abstracts} of the 2021 {CHI} {Conference} on {Human} {Factors} in {Computing} {Systems}}} \emph{(\bibinfo{series}{{CHI} {EA} '21})}. \bibinfo{publisher}{Association for Computing Machinery}, \bibinfo{address}{New York, NY, USA}, \bibinfo{pages}{1--6}.
\newblock
\showISBNx{978-1-4503-8095-9}
\urldef\tempurl%
\url{https://doi.org/10.1145/3411763.3441346}
\showDOI{\tempurl}


\bibitem[Li et~al\mbox{.}(2023)]%
        {li_we_2023}
\bibfield{author}{\bibinfo{person}{Lingyuan Li}, \bibinfo{person}{Guo Freeman}, \bibinfo{person}{Kelsea Schulenberg}, {and} \bibinfo{person}{Dane Acena}.} \bibinfo{year}{2023}\natexlab{}.
\newblock \showarticletitle{"{We} {Cried} on {Each} {Other}’s {Shoulders}": {How} {LGBTQ}+ {Individuals} {Experience} {Social} {Support} in {Social} {Virtual} {Reality}}. In \bibinfo{booktitle}{\emph{Proceedings of the 2023 {CHI} {Conference} on {Human} {Factors} in {Computing} {Systems}}} \emph{(\bibinfo{series}{{CHI} '23})}. \bibinfo{publisher}{Association for Computing Machinery}, \bibinfo{address}{New York, NY, USA}, \bibinfo{pages}{1--16}.
\newblock
\showISBNx{978-1-4503-9421-5}
\urldef\tempurl%
\url{https://doi.org/10.1145/3544548.3581530}
\showDOI{\tempurl}


\bibitem[Litt and Hargittai(2014)]%
        {litt_bumpy_2014}
\bibfield{author}{\bibinfo{person}{Eden Litt} {and} \bibinfo{person}{Eszter Hargittai}.} \bibinfo{year}{2014}\natexlab{}.
\newblock \showarticletitle{A bumpy ride on the information superhighway: {Exploring} turbulence online}.
\newblock \bibinfo{journal}{\emph{Computers in Human Behavior}}  \bibinfo{volume}{36} (\bibinfo{date}{July} \bibinfo{year}{2014}), \bibinfo{pages}{520--529}.
\newblock
\showISSN{0747-5632}
\urldef\tempurl%
\url{https://doi.org/10.1016/j.chb.2014.04.027}
\showDOI{\tempurl}


\bibitem[McInroy(2020)]%
        {mcinroy_building_2020}
\bibfield{author}{\bibinfo{person}{Lauren~B. McInroy}.} \bibinfo{year}{2020}\natexlab{}.
\newblock \showarticletitle{Building connections and slaying basilisks: fostering support, resilience, and positive adjustment for sexual and gender minority youth in online fandom communities}.
\newblock \bibinfo{journal}{\emph{Information, Communication \& Society}} \bibinfo{volume}{23}, \bibinfo{number}{13} (\bibinfo{date}{Nov.} \bibinfo{year}{2020}), \bibinfo{pages}{1874--1891}.
\newblock
\showISSN{1369-118X}
\urldef\tempurl%
\url{https://doi.org/10.1080/1369118X.2019.1623902}
\showDOI{\tempurl}
\newblock
\shownote{Publisher: Routledge \_eprint: https://doi.org/10.1080/1369118X.2019.1623902}.


\bibitem[Moore et~al\mbox{.}(2009)]%
        {moore_3d_2009}
\bibfield{author}{\bibinfo{person}{Robert Moore}, \bibinfo{person}{E. Hankinson~Gathman}, {and} \bibinfo{person}{Nicolas Ducheneaut}.} \bibinfo{year}{2009}\natexlab{}.
\newblock \showarticletitle{From {3D} {Space} to {Third} {Place}: {The} {Social} {Life} of {Small} {Virtual} {Spaces}}.
\newblock \bibinfo{journal}{\emph{Human Organization}} \bibinfo{volume}{68}, \bibinfo{number}{2} (\bibinfo{date}{May} \bibinfo{year}{2009}), \bibinfo{pages}{230--240}.
\newblock
\showISSN{0018-7259}
\urldef\tempurl%
\url{https://doi.org/10.17730/humo.68.2.q673k16185u68v15}
\showDOI{\tempurl}


\bibitem[Morahan-Martin and Schumacher(2003)]%
        {morahan-martin_loneliness_2003}
\bibfield{author}{\bibinfo{person}{Janet Morahan-Martin} {and} \bibinfo{person}{Phyllis Schumacher}.} \bibinfo{year}{2003}\natexlab{}.
\newblock \showarticletitle{Loneliness and social uses of the {Internet}}.
\newblock \bibinfo{journal}{\emph{Computers in Human Behavior}} \bibinfo{volume}{19}, \bibinfo{number}{6} (\bibinfo{date}{Nov.} \bibinfo{year}{2003}), \bibinfo{pages}{659--671}.
\newblock
\showISSN{0747-5632}
\urldef\tempurl%
\url{https://doi.org/10.1016/S0747-5632(03)00040-2}
\showDOI{\tempurl}


\bibitem[Nardi(2010)]%
        {nardi_my_2010}
\bibfield{author}{\bibinfo{person}{Bonnie Nardi}.} \bibinfo{year}{2010}\natexlab{}.
\newblock \bibinfo{booktitle}{\emph{My {Life} as a {Night} {Elf} {Priest}: {An} {Anthropological} {Account} of {World} of {Warcraft}}}.
\newblock \bibinfo{publisher}{University of Michigan Press}.
\newblock
\showISBNx{978-0-472-07098-5 978-0-472-05098-7 978-0-472-90043-5}
\urldef\tempurl%
\url{https://doi.org/10.3998/toi.8008655.0001.001}
\showDOI{\tempurl}
\newblock
\shownote{Accepted: 2019-11-09 03:00:32}.


\bibitem[Nardi(2005)]%
        {nardi_beyond_2005}
\bibfield{author}{\bibinfo{person}{Bonnie~A. Nardi}.} \bibinfo{year}{2005}\natexlab{}.
\newblock \showarticletitle{Beyond {Bandwidth}: {Dimensions} of {Connection} in {Interpersonal} {Communication}}.
\newblock \bibinfo{journal}{\emph{Computer Supported Cooperative Work (CSCW)}} \bibinfo{volume}{14}, \bibinfo{number}{2} (\bibinfo{date}{April} \bibinfo{year}{2005}), \bibinfo{pages}{91--130}.
\newblock
\showISSN{1573-7551}
\urldef\tempurl%
\url{https://doi.org/10.1007/s10606-004-8127-9}
\showDOI{\tempurl}


\bibitem[Nguyen et~al\mbox{.}(2022)]%
        {nguyen_staying_2022}
\bibfield{author}{\bibinfo{person}{Minh~Hao Nguyen}, \bibinfo{person}{Jonathan Gruber}, \bibinfo{person}{Will Marler}, \bibinfo{person}{Amanda Hunsaker}, \bibinfo{person}{Jaelle Fuchs}, {and} \bibinfo{person}{Eszter Hargittai}.} \bibinfo{year}{2022}\natexlab{}.
\newblock \showarticletitle{Staying connected while physically apart: {Digital} communication when face-to-face interactions are limited}.
\newblock \bibinfo{journal}{\emph{New Media \& Society}} \bibinfo{volume}{24}, \bibinfo{number}{9} (\bibinfo{date}{Sept.} \bibinfo{year}{2022}), \bibinfo{pages}{2046--2067}.
\newblock
\showISSN{1461-4448}
\urldef\tempurl%
\url{https://doi.org/10.1177/1461444820985442}
\showDOI{\tempurl}
\newblock
\shownote{Publisher: SAGE Publications}.


\bibitem[Norton et~al\mbox{.}(2007)]%
        {norton_less_2007}
\bibfield{author}{\bibinfo{person}{Michael~I. Norton}, \bibinfo{person}{Jeana~H. Frost}, {and} \bibinfo{person}{Dan Ariely}.} \bibinfo{year}{2007}\natexlab{}.
\newblock \showarticletitle{Less is more: {The} lure of ambiguity, or why familiarity breeds contempt}.
\newblock \bibinfo{journal}{\emph{Journal of Personality and Social Psychology}} \bibinfo{volume}{92}, \bibinfo{number}{1} (\bibinfo{year}{2007}), \bibinfo{pages}{97--105}.
\newblock
\showISSN{1939-1315}
\urldef\tempurl%
\url{https://doi.org/10.1037/0022-3514.92.1.97}
\showDOI{\tempurl}
\newblock
\shownote{Place: US Publisher: American Psychological Association}.


\bibitem[Oswald et~al\mbox{.}(2004)]%
        {oswald_friendship_2004}
\bibfield{author}{\bibinfo{person}{Debra~L. Oswald}, \bibinfo{person}{Eddie~M. Clark}, {and} \bibinfo{person}{Cheryl~M. Kelly}.} \bibinfo{year}{2004}\natexlab{}.
\newblock \showarticletitle{Friendship {Maintenance}: {An} {Analysis} of {Individual} and {Dyad} {Behaviors}}.
\newblock \bibinfo{journal}{\emph{Journal of Social and Clinical Psychology}} \bibinfo{volume}{23}, \bibinfo{number}{3} (\bibinfo{date}{June} \bibinfo{year}{2004}), \bibinfo{pages}{413--441}.
\newblock
\showISSN{0736-7236}
\urldef\tempurl%
\url{https://doi.org/10.1521/jscp.23.3.413.35460}
\showDOI{\tempurl}
\newblock
\shownote{Publisher: Guilford Publications Inc.}.


\bibitem[Patulny and Bower(2022)]%
        {patulny_beware_2022}
\bibfield{author}{\bibinfo{person}{Roger Patulny} {and} \bibinfo{person}{Marlee Bower}.} \bibinfo{year}{2022}\natexlab{}.
\newblock \showarticletitle{Beware the “loneliness gap”? {Examining} emerging inequalities and long-term risks of loneliness and isolation emerging from {COVID}-19}.
\newblock \bibinfo{journal}{\emph{Australian Journal of Social Issues}} \bibinfo{volume}{57}, \bibinfo{number}{3} (\bibinfo{year}{2022}), \bibinfo{pages}{562--583}.
\newblock
\showISSN{1839-4655}
\urldef\tempurl%
\url{https://doi.org/10.1002/ajs4.223}
\showDOI{\tempurl}
\newblock
\shownote{\_eprint: https://onlinelibrary.wiley.com/doi/pdf/10.1002/ajs4.223}.


\bibitem[Prescott et~al\mbox{.}(2017)]%
        {prescott_peer_2017}
\bibfield{author}{\bibinfo{person}{Julie Prescott}, \bibinfo{person}{Terry Hanley}, {and} \bibinfo{person}{Katalin Ujhelyi}.} \bibinfo{year}{2017}\natexlab{}.
\newblock \showarticletitle{Peer {Communication} in {Online} {Mental} {Health} {Forums} for {Young} {People}: {Directional} and {Nondirectional} {Support}}.
\newblock \bibinfo{journal}{\emph{JMIR Mental Health}} \bibinfo{volume}{4}, \bibinfo{number}{3} (\bibinfo{date}{Aug.} \bibinfo{year}{2017}), \bibinfo{pages}{e6921}.
\newblock
\urldef\tempurl%
\url{https://doi.org/10.2196/mental.6921}
\showDOI{\tempurl}
\newblock
\shownote{Company: JMIR Mental Health Distributor: JMIR Mental Health Institution: JMIR Mental Health Label: JMIR Mental Health Publisher: JMIR Publications Inc., Toronto, Canada}.


\bibitem[Ramirez(2018)]%
        {sukthankar_good_2018}
\bibfield{author}{\bibinfo{person}{Fanny~Anne Ramirez}.} \bibinfo{year}{2018}\natexlab{}.
\newblock \showarticletitle{From {Good} {Associates} to {True} {Friends}: {An} {Exploration} of {Friendship} {Practices} in {Massively} {Multiplayer} {Online} {Games}}.
\newblock In \bibinfo{booktitle}{\emph{Social {Interactions} in {Virtual} {Worlds}: {An} {Interdisciplinary} {Perspective}}}, \bibfield{editor}{\bibinfo{person}{Gita Sukthankar}, \bibinfo{person}{Kiran Lakkaraju}, {and} \bibinfo{person}{Rolf~T. Wigand}} (Eds.). \bibinfo{publisher}{Cambridge University Press}, \bibinfo{address}{Cambridge}, \bibinfo{pages}{62--79}.
\newblock
\showISBNx{978-1-107-12882-8}
\urldef\tempurl%
\url{https://doi.org/10.1017/9781316422823.004}
\showDOI{\tempurl}


\bibitem[Ren et~al\mbox{.}(2012)]%
        {ren_building_2012}
\bibfield{author}{\bibinfo{person}{Yuqing Ren}, \bibinfo{person}{F.~Maxwell Harper}, \bibinfo{person}{Sara Drenner}, \bibinfo{person}{Loren Terveen}, \bibinfo{person}{Sara Kiesler}, \bibinfo{person}{John Riedl}, {and} \bibinfo{person}{Robert~E. Kraut}.} \bibinfo{year}{2012}\natexlab{}.
\newblock \showarticletitle{Building {Member} {Attachment} in {Online} {Communities}: {Applying} {Theories} of {Group} {Identity} and {Interpersonal} {Bonds}}.
\newblock \bibinfo{journal}{\emph{MIS Quarterly}} \bibinfo{volume}{36}, \bibinfo{number}{3} (\bibinfo{year}{2012}), \bibinfo{pages}{841--864}.
\newblock
\showISSN{0276-7783}
\urldef\tempurl%
\url{https://doi.org/10.2307/41703483}
\showDOI{\tempurl}
\newblock
\shownote{Publisher: Management Information Systems Research Center, University of Minnesota}.


\bibitem[Ringland(2019)]%
        {ringland_place_2019}
\bibfield{author}{\bibinfo{person}{Kathryn~E. Ringland}.} \bibinfo{year}{2019}\natexlab{}.
\newblock \showarticletitle{A {Place} to {Play}: {The} ({Dis}){Abled} {Embodied} {Experience} for {Autistic} {Children} in {Online} {Spaces}}. In \bibinfo{booktitle}{\emph{Proceedings of the 2019 {CHI} {Conference} on {Human} {Factors} in {Computing} {Systems}}} \emph{(\bibinfo{series}{{CHI} '19})}. \bibinfo{publisher}{Association for Computing Machinery}, \bibinfo{address}{New York, NY, USA}, \bibinfo{pages}{1--14}.
\newblock
\showISBNx{978-1-4503-5970-2}
\urldef\tempurl%
\url{https://doi.org/10.1145/3290605.3300518}
\showDOI{\tempurl}


\bibitem[Ringland et~al\mbox{.}(2016)]%
        {ringland_will_2016}
\bibfield{author}{\bibinfo{person}{Kathryn~E. Ringland}, \bibinfo{person}{Christine~T. Wolf}, \bibinfo{person}{Heather Faucett}, \bibinfo{person}{Lynn Dombrowski}, {and} \bibinfo{person}{Gillian~R. Hayes}.} \bibinfo{year}{2016}\natexlab{}.
\newblock \showarticletitle{"{Will} {I} always be not social?": {Re}-{Conceptualizing} {Sociality} in the {Context} of a {Minecraft} {Community} for {Autism}}. In \bibinfo{booktitle}{\emph{Proceedings of the 2016 {CHI} {Conference} on {Human} {Factors} in {Computing} {Systems}}} \emph{(\bibinfo{series}{{CHI} '16})}. \bibinfo{publisher}{Association for Computing Machinery}, \bibinfo{address}{New York, NY, USA}, \bibinfo{pages}{1256--1269}.
\newblock
\showISBNx{978-1-4503-3362-7}
\urldef\tempurl%
\url{https://doi.org/10.1145/2858036.2858038}
\showDOI{\tempurl}


\bibitem[Rubin et~al\mbox{.}(2009)]%
        {rubin_handbook_2009}
\bibfield{editor}{\bibinfo{person}{Kenneth~H. Rubin}, \bibinfo{person}{William~M. Bukowski}, {and} \bibinfo{person}{Brett Laursen}} (Eds.). \bibinfo{year}{2009}\natexlab{}.
\newblock \bibinfo{booktitle}{\emph{Handbook of peer interactions, relationships, and groups}}.
\newblock \bibinfo{publisher}{The Guilford Press}, \bibinfo{address}{New York, NY, US}.
\newblock
\showISBNx{978-1-59385-441-6}
\newblock
\shownote{Pages: xvii, 654}.


\bibitem[Sassenberg(2002)]%
        {sassenberg_common_2002}
\bibfield{author}{\bibinfo{person}{Kai Sassenberg}.} \bibinfo{year}{2002}\natexlab{}.
\newblock \showarticletitle{Common bond and common identity groups on the {Internet}: {Attachment} and normative behavior in on-topic and off-topic chats}.
\newblock \bibinfo{journal}{\emph{Group Dynamics: Theory, Research, and Practice}} \bibinfo{volume}{6}, \bibinfo{number}{1} (\bibinfo{year}{2002}), \bibinfo{pages}{27--37}.
\newblock
\showISSN{1930-7802}
\urldef\tempurl%
\url{https://doi.org/10.1037/1089-2699.6.1.27}
\showDOI{\tempurl}
\newblock
\shownote{Place: US Publisher: Educational Publishing Foundation}.


\bibitem[Scott et~al\mbox{.}(2022)]%
        {scott_social_2022}
\bibfield{author}{\bibinfo{person}{Riley~A. Scott}, \bibinfo{person}{Jaimee Stuart}, \bibinfo{person}{Bonnie~L. Barber}, \bibinfo{person}{Karlee~J. O'Donnell}, {and} \bibinfo{person}{Alexander~W. O'Donnell}.} \bibinfo{year}{2022}\natexlab{}.
\newblock \showarticletitle{Social connections during physical isolation: {How} a shift to online interaction explains friendship satisfaction and social well-being}.
\newblock \bibinfo{journal}{\emph{Cyberpsychology: Journal of Psychosocial Research on Cyberspace}} \bibinfo{volume}{16}, \bibinfo{number}{2} (\bibinfo{date}{April} \bibinfo{year}{2022}).
\newblock
\showISSN{1802-7962}
\urldef\tempurl%
\url{https://doi.org/10.5817/CP2022-2-10}
\showDOI{\tempurl}
\newblock
\shownote{Number: 2}.


\bibitem[Seaman(2008)]%
        {seaman_adopting_2008}
\bibfield{author}{\bibinfo{person}{Jayson Seaman}.} \bibinfo{year}{2008}\natexlab{}.
\newblock \showarticletitle{Adopting a {Grounded} {Theory} {Approach} to {Cultural}-{Historical} {Research}: {Conflicting} {Methodologies} or {Complementary} {Methods}?}
\newblock \bibinfo{journal}{\emph{International Journal of Qualitative Methods}} \bibinfo{volume}{7}, \bibinfo{number}{1} (\bibinfo{date}{March} \bibinfo{year}{2008}), \bibinfo{pages}{1--17}.
\newblock
\showISSN{1609-4069}
\urldef\tempurl%
\url{https://doi.org/10.1177/160940690800700101}
\showDOI{\tempurl}
\newblock
\shownote{Publisher: SAGE Publications Inc}.


\bibitem[Sharma et~al\mbox{.}(2023)]%
        {sharma_humanai_2023}
\bibfield{author}{\bibinfo{person}{Ashish Sharma}, \bibinfo{person}{Inna~W. Lin}, \bibinfo{person}{Adam~S. Miner}, \bibinfo{person}{David~C. Atkins}, {and} \bibinfo{person}{Tim Althoff}.} \bibinfo{year}{2023}\natexlab{}.
\newblock \showarticletitle{Human–{AI} collaboration enables more empathic conversations in text-based peer-to-peer mental health support}.
\newblock \bibinfo{journal}{\emph{Nature Machine Intelligence}} \bibinfo{volume}{5}, \bibinfo{number}{1} (\bibinfo{date}{Jan.} \bibinfo{year}{2023}), \bibinfo{pages}{46--57}.
\newblock
\showISSN{2522-5839}
\urldef\tempurl%
\url{https://doi.org/10.1038/s42256-022-00593-2}
\showDOI{\tempurl}
\newblock
\shownote{Publisher: Nature Publishing Group}.


\bibitem[Sheng and Kairam(2020)]%
        {sheng_virtual_2020}
\bibfield{author}{\bibinfo{person}{Jeff~T. Sheng} {and} \bibinfo{person}{Sanjay~R. Kairam}.} \bibinfo{year}{2020}\natexlab{}.
\newblock \showarticletitle{From {Virtual} {Strangers} to {IRL} {Friends}: {Relationship} {Development} in {Livestreaming} {Communities} on {Twitch}}.
\newblock \bibinfo{journal}{\emph{Proceedings of the ACM on Human-Computer Interaction}} \bibinfo{volume}{4}, \bibinfo{number}{CSCW2} (\bibinfo{date}{Oct.} \bibinfo{year}{2020}), \bibinfo{pages}{94:1--94:34}.
\newblock
\urldef\tempurl%
\url{https://doi.org/10.1145/3415165}
\showDOI{\tempurl}


\bibitem[Steinkuehler and Williams(2006)]%
        {steinkuehler_where_2006}
\bibfield{author}{\bibinfo{person}{Constance~A. Steinkuehler} {and} \bibinfo{person}{Dmitri Williams}.} \bibinfo{year}{2006}\natexlab{}.
\newblock \showarticletitle{Where {Everybody} {Knows} {Your} ({Screen}) {Name}: {Online} {Games} as “{Third} {Places}”}.
\newblock \bibinfo{journal}{\emph{Journal of Computer-Mediated Communication}} \bibinfo{volume}{11}, \bibinfo{number}{4} (\bibinfo{date}{July} \bibinfo{year}{2006}), \bibinfo{pages}{885--909}.
\newblock
\showISSN{1083-6101}
\urldef\tempurl%
\url{https://doi.org/10.1111/j.1083-6101.2006.00300.x}
\showDOI{\tempurl}


\bibitem[Stuart et~al\mbox{.}(2021)]%
        {stuart_online_2021}
\bibfield{author}{\bibinfo{person}{Jaimee Stuart}, \bibinfo{person}{Karlee O'Donnell}, \bibinfo{person}{Alex O'Donnell}, \bibinfo{person}{Riley Scott}, {and} \bibinfo{person}{Bonnie Barber}.} \bibinfo{year}{2021}\natexlab{}.
\newblock \showarticletitle{Online {Social} {Connection} as a {Buffer} of {Health} {Anxiety} and {Isolation} {During} {COVID}-19}.
\newblock \bibinfo{journal}{\emph{Cyberpsychology, Behavior, and Social Networking}} \bibinfo{volume}{24}, \bibinfo{number}{8} (\bibinfo{date}{Aug.} \bibinfo{year}{2021}), \bibinfo{pages}{521--525}.
\newblock
\showISSN{2152-2715}
\urldef\tempurl%
\url{https://doi.org/10.1089/cyber.2020.0645}
\showDOI{\tempurl}
\newblock
\shownote{Publisher: Mary Ann Liebert, Inc., publishers}.


\bibitem[Tang(2010)]%
        {tang_development_2010}
\bibfield{author}{\bibinfo{person}{Lijun Tang}.} \bibinfo{year}{2010}\natexlab{}.
\newblock \showarticletitle{Development of {Online} {Friendship} in {Different} {Social} {Spaces}}.
\newblock \bibinfo{journal}{\emph{Information, Communication \& Society}} \bibinfo{volume}{13}, \bibinfo{number}{4} (\bibinfo{date}{June} \bibinfo{year}{2010}), \bibinfo{pages}{615--633}.
\newblock
\showISSN{1369-118X}
\urldef\tempurl%
\url{https://doi.org/10.1080/13691180902998639}
\showDOI{\tempurl}
\newblock
\shownote{Publisher: Routledge \_eprint: https://doi.org/10.1080/13691180902998639}.


\bibitem[Tian(2013)]%
        {tian_social_2013}
\bibfield{author}{\bibinfo{person}{Qing Tian}.} \bibinfo{year}{2013}\natexlab{}.
\newblock \showarticletitle{Social {Anxiety}, {Motivation}, {Self}-{Disclosure}, and {Computer}-{Mediated} {Friendship}: {A} {Path} {Analysis} of the {Social} {Interaction} in the {Blogosphere}}.
\newblock \bibinfo{journal}{\emph{Communication Research}} \bibinfo{volume}{40}, \bibinfo{number}{2} (\bibinfo{date}{April} \bibinfo{year}{2013}), \bibinfo{pages}{237--260}.
\newblock
\showISSN{0093-6502}
\urldef\tempurl%
\url{https://doi.org/10.1177/0093650211420137}
\showDOI{\tempurl}
\newblock
\shownote{Publisher: SAGE Publications Inc}.


\bibitem[Towner et~al\mbox{.}(2022)]%
        {towner_virtual_2022}
\bibfield{author}{\bibinfo{person}{Emily Towner}, \bibinfo{person}{Livia Tomova}, \bibinfo{person}{Danielle Ladensack}, \bibinfo{person}{Kristen Chu}, {and} \bibinfo{person}{Bridget Callaghan}.} \bibinfo{year}{2022}\natexlab{}.
\newblock \showarticletitle{Virtual social interaction and loneliness among emerging adults amid the {COVID}-19 pandemic}.
\newblock \bibinfo{journal}{\emph{Current Research in Ecological and Social Psychology}}  \bibinfo{volume}{3} (\bibinfo{date}{Jan.} \bibinfo{year}{2022}), \bibinfo{pages}{100058}.
\newblock
\showISSN{2666-6227}
\urldef\tempurl%
\url{https://doi.org/10.1016/j.cresp.2022.100058}
\showDOI{\tempurl}


\bibitem[Utz(2000)]%
        {utz_social_2000}
\bibfield{author}{\bibinfo{person}{Sonja Utz}.} \bibinfo{year}{2000}\natexlab{}.
\newblock \showarticletitle{Social information processing in {MUDs}: {The} development of friendships in virtual worlds}.
\newblock \bibinfo{journal}{\emph{Journal of online behavior}} \bibinfo{volume}{1}, \bibinfo{number}{1} (\bibinfo{year}{2000}), \bibinfo{pages}{2002}.
\newblock
\urldef\tempurl%
\url{https://www.researchgate.net/profile/Sonja-Utz/publication/344749824_Social_Information_Processing_in_MUDs_The_development_of_friendships_in_virtual_worlds/links/5f8d8cfaa6fdccfd7b6c1a45/Social-Information-Processing-in-MUDs-The-development-of-friendships-in-virtual-worlds.pdf?_sg%5B0%5D=started_experiment_milestone&origin=journalDetail}
\showURL{%
\tempurl}


\bibitem[Vetere et~al\mbox{.}(2005)]%
        {vetere_mediating_2005}
\bibfield{author}{\bibinfo{person}{Frank Vetere}, \bibinfo{person}{Martin~R. Gibbs}, \bibinfo{person}{Jesper Kjeldskov}, \bibinfo{person}{Steve Howard}, \bibinfo{person}{Florian~'Floyd' Mueller}, \bibinfo{person}{Sonja Pedell}, \bibinfo{person}{Karen Mecoles}, {and} \bibinfo{person}{Marcus Bunyan}.} \bibinfo{year}{2005}\natexlab{}.
\newblock \showarticletitle{Mediating intimacy: designing technologies to support strong-tie relationships}. In \bibinfo{booktitle}{\emph{Proceedings of the {SIGCHI} {Conference} on {Human} {Factors} in {Computing} {Systems}}} \emph{(\bibinfo{series}{{CHI} '05})}. \bibinfo{publisher}{Association for Computing Machinery}, \bibinfo{address}{New York, NY, USA}, \bibinfo{pages}{471--480}.
\newblock
\showISBNx{978-1-58113-998-3}
\urldef\tempurl%
\url{https://doi.org/10.1145/1054972.1055038}
\showDOI{\tempurl}


\bibitem[Walther(2008)]%
        {walther_social_2008}
\bibfield{author}{\bibinfo{person}{Joseph Walther}.} \bibinfo{year}{2008}\natexlab{}.
\newblock \showarticletitle{Social {Information} {Processing} {Theory}: {Impressions} and {Relationship} {Development} {Online}}.
\newblock In \bibinfo{booktitle}{\emph{Engaging {Theories} in {Interpersonal} {Communication}: {Multiple} {Perspectives}}}. \bibinfo{publisher}{SAGE Publications}, \bibinfo{address}{London}, \bibinfo{pages}{391--404}.
\newblock
\showISBNx{978-1-4129-3852-5}
\urldef\tempurl%
\url{https://doi.org/10.4135/9781483329529.n29}
\showDOI{\tempurl}


\bibitem[WALTHER(1992)]%
        {walther_interpersonal_1992}
\bibfield{author}{\bibinfo{person}{JOSEPH~B. WALTHER}.} \bibinfo{year}{1992}\natexlab{}.
\newblock \showarticletitle{Interpersonal {Effects} in {Computer}-{Mediated} {Interaction}: {A} {Relational} {Perspective}}.
\newblock \bibinfo{journal}{\emph{Communication Research}} \bibinfo{volume}{19}, \bibinfo{number}{1} (\bibinfo{date}{Feb.} \bibinfo{year}{1992}), \bibinfo{pages}{52--90}.
\newblock
\showISSN{0093-6502}
\urldef\tempurl%
\url{https://doi.org/10.1177/009365092019001003}
\showDOI{\tempurl}
\newblock
\shownote{Publisher: SAGE Publications Inc}.


\bibitem[Walther(1996)]%
        {walther_computer-mediated_1996}
\bibfield{author}{\bibinfo{person}{Joseph~B. Walther}.} \bibinfo{year}{1996}\natexlab{}.
\newblock \showarticletitle{Computer-mediated communication: {Impersonal}, interpersonal, and hyperpersonal interaction}.
\newblock \bibinfo{journal}{\emph{Communication Research}} \bibinfo{volume}{23}, \bibinfo{number}{1} (\bibinfo{year}{1996}), \bibinfo{pages}{3--43}.
\newblock
\showISSN{1552-3810}
\urldef\tempurl%
\url{https://doi.org/10.1177/009365096023001001}
\showDOI{\tempurl}
\newblock
\shownote{Place: US Publisher: Sage Publications}.


\bibitem[Wang et~al\mbox{.}(2020)]%
        {wang_again_2020}
\bibfield{author}{\bibinfo{person}{Cheng~Yao Wang}, \bibinfo{person}{Mose Sakashita}, \bibinfo{person}{Upol Ehsan}, \bibinfo{person}{Jingjin Li}, {and} \bibinfo{person}{Andrea~Stevenson Won}.} \bibinfo{year}{2020}\natexlab{}.
\newblock \showarticletitle{Again, {Together}: {Socially} {Reliving} {Virtual} {Reality} {Experiences} {When} {Separated}}. In \bibinfo{booktitle}{\emph{Proceedings of the 2020 {CHI} {Conference} on {Human} {Factors} in {Computing} {Systems}}} \emph{(\bibinfo{series}{{CHI} '20})}. \bibinfo{publisher}{Association for Computing Machinery}, \bibinfo{address}{New York, NY, USA}, \bibinfo{pages}{1--12}.
\newblock
\showISBNx{978-1-4503-6708-0}
\urldef\tempurl%
\url{https://doi.org/10.1145/3313831.3376642}
\showDOI{\tempurl}


\bibitem[Whitty and Gavin(2001)]%
        {whitty_agesexlocation_2001}
\bibfield{author}{\bibinfo{person}{M. Whitty} {and} \bibinfo{person}{J. Gavin}.} \bibinfo{year}{2001}\natexlab{}.
\newblock \showarticletitle{Age/sex/location: uncovering the social cues in the development of online relationships}.
\newblock \bibinfo{journal}{\emph{Cyberpsychology \& Behavior: The Impact of the Internet, Multimedia and Virtual Reality on Behavior and Society}} \bibinfo{volume}{4}, \bibinfo{number}{5} (\bibinfo{date}{Oct.} \bibinfo{year}{2001}), \bibinfo{pages}{623--630}.
\newblock
\showISSN{1094-9313}
\urldef\tempurl%
\url{https://doi.org/10.1089/109493101753235223}
\showDOI{\tempurl}


\bibitem[Wisniewski et~al\mbox{.}(2017a)]%
        {wisniewski_parental_2017}
\bibfield{author}{\bibinfo{person}{Pamela Wisniewski}, \bibinfo{person}{Arup~Kumar Ghosh}, \bibinfo{person}{Heng Xu}, \bibinfo{person}{Mary~Beth Rosson}, {and} \bibinfo{person}{John~M. Carroll}.} \bibinfo{year}{2017}\natexlab{a}.
\newblock \showarticletitle{Parental {Control} vs. {Teen} {Self}-{Regulation}: {Is} there a middle ground for mobile online safety?}. In \bibinfo{booktitle}{\emph{Proceedings of the 2017 {ACM} {Conference} on {Computer} {Supported} {Cooperative} {Work} and {Social} {Computing}}} \emph{(\bibinfo{series}{{CSCW} '17})}. \bibinfo{publisher}{Association for Computing Machinery}, \bibinfo{address}{New York, NY, USA}, \bibinfo{pages}{51--69}.
\newblock
\showISBNx{978-1-4503-4335-0}
\urldef\tempurl%
\url{https://doi.org/10.1145/2998181.2998352}
\showDOI{\tempurl}


\bibitem[Wisniewski et~al\mbox{.}(2017b)]%
        {wisniewski_parents_2017}
\bibfield{author}{\bibinfo{person}{Pamela Wisniewski}, \bibinfo{person}{Heng Xu}, \bibinfo{person}{Mary~Beth Rosson}, {and} \bibinfo{person}{John~M. Carroll}.} \bibinfo{year}{2017}\natexlab{b}.
\newblock \showarticletitle{Parents {Just} {Don}'t {Understand}: {Why} {Teens} {Don}'t {Talk} to {Parents} about {Their} {Online} {Risk} {Experiences}}. In \bibinfo{booktitle}{\emph{Proceedings of the 2017 {ACM} {Conference} on {Computer} {Supported} {Cooperative} {Work} and {Social} {Computing}}} \emph{(\bibinfo{series}{{CSCW} '17})}. \bibinfo{publisher}{Association for Computing Machinery}, \bibinfo{address}{New York, NY, USA}, \bibinfo{pages}{523--540}.
\newblock
\showISBNx{978-1-4503-4335-0}
\urldef\tempurl%
\url{https://doi.org/10.1145/2998181.2998236}
\showDOI{\tempurl}


\bibitem[Yum and Hara(2005)]%
        {yum_computer-mediated_2005}
\bibfield{author}{\bibinfo{person}{Young-ok Yum} {and} \bibinfo{person}{Kazuya Hara}.} \bibinfo{year}{2005}\natexlab{}.
\newblock \showarticletitle{Computer-{Mediated} {Relationship} {Development}: a {Cross}-{Cultural} {Comparison}}.
\newblock \bibinfo{journal}{\emph{Journal of Computer-Mediated Communication}} \bibinfo{volume}{11}, \bibinfo{number}{1} (\bibinfo{date}{Nov.} \bibinfo{year}{2005}), \bibinfo{pages}{133--152}.
\newblock
\showISSN{1083-6101}
\urldef\tempurl%
\url{https://doi.org/10.1111/j.1083-6101.2006.tb00307.x}
\showDOI{\tempurl}


\bibitem[Zhu et~al\mbox{.}(2014)]%
        {zhu_impact_2014}
\bibfield{author}{\bibinfo{person}{Haiyi Zhu}, \bibinfo{person}{Robert~E. Kraut}, {and} \bibinfo{person}{Aniket Kittur}.} \bibinfo{year}{2014}\natexlab{}.
\newblock \showarticletitle{The impact of membership overlap on the survival of online communities}. In \bibinfo{booktitle}{\emph{Proceedings of the {SIGCHI} {Conference} on {Human} {Factors} in {Computing} {Systems}}} \emph{(\bibinfo{series}{{CHI} '14})}. \bibinfo{publisher}{Association for Computing Machinery}, \bibinfo{address}{New York, NY, USA}, \bibinfo{pages}{281--290}.
\newblock
\showISBNx{978-1-4503-2473-1}
\urldef\tempurl%
\url{https://doi.org/10.1145/2556288.2557213}
\showDOI{\tempurl}


\end{thebibliography}

\end{document}